\documentclass[a4paper,fleqn,usenatbib]{mnras}

\usepackage{newtxtext,newtxmath}

\usepackage[T1]{fontenc}
\usepackage{ae,aecompl}

\usepackage{graphicx}	
\usepackage{amsmath}	
\usepackage{amssymb}	
\usepackage{caption}
\pdfoutput=1

\title[Faraday effects and Sgr A* polarization.]{The impact of
  Faraday effects on polarized black hole images of Sagittarius A*.}

\author[A. Jim\'enez-Rosales et al.]{
Alejandra Jim\'enez-Rosales\thanks{E-mail: ajimenez@mpe.mpg.de}
and Jason Dexter\thanks{E-mail: jdexter@mpe.mpg.de}
\\
Max-Plack-Institut f\"ur Extraterrestrische Physik, Giessenbachstr. 1, 85748 Garching, Germany\\
}

\date{Accepted XXX. Received YYY; in original form ZZZ}

\pubyear{2017}

\begin{document}
\label{firstpage}
\pagerange{\pageref{firstpage}--\pageref{lastpage}}
\maketitle

\begin{abstract}
We study model images and polarization maps of Sagittarius A* at 230 GHz. 
We post-process GRMHD simulations and perform a fully relativistic radiative transfer calculation of the 
emitted synchrotron radiation to obtain polarized images for a range
of mass accretion rates and electron temperatures. At low accretion
rates, the polarization map traces the underlying toroidal magnetic
field geometry. At high accretion rates, we find that Faraday rotation
internal to the emission region can depolarize and scramble the map. We measure the net linear polarization
fraction and find that high accretion rate ``jet-disc'' models are heavily depolarized and are therefore disfavoured.
We show how Event Horizon Telescope measurements of the
polarized ``correlation length'' over the image provide a
model-independent upper limit on the strength of these Faraday
effects, and constrain plasma properties like the electron 
temperature and magnetic field strength. 

\end{abstract}



\section{Introduction}

The compact radio source Sagittarius A* (Sgr A*) is the closest
supermassive black hole candidate to Earth \citep[e.g.,][]{Genzel+2010,Falcke+2013}. 
With a mass $M\sim4.3\times10^{6}$ solar masses 
and at a distance $D\sim8.3$ kpc \citep{Boehle+2016,Gillessen+2017}, Sgr A* is the black hole with 
the largest apparent angular size on the sky (with a shadow of $\sim50\mu$as\footnote{The angular 
size of Sgr A* is given by $R_{\rm{s}}/D\sim10\ \mu$as, where $R_{\rm{s}}\sim1.36\times10^{12}\ \rm{cm}$ is 
the Schwarzschild radius. }), which makes it an excellent laboratory for studying 
accretion physics around black holes and for probing general relativistic effects. 
Sgr A* emits most of its luminosity from synchrotron radiation in 
what is called the ``submillimetre bump.'' At these wavelengths
($\lesssim 1$ mm), the
radiation is expected to be optically thin and originate from close to
the black hole \citep{Falcke+2000,Bower+2015}. 

Models of radiatively inefficient accretion flows
\citep[RIAFs][]{Narayan+1995,Quataert+1999,Yuan+2003} and magnetised
jets \citep{Falcke+2000,Yuan+2002} have been developed to explain the
radio spectrum of Sgr A*. Such models can now be realised using general relativistic 
magnetohydrodynamic (GRMHD) simulations, which capture the
time-dependent accretion process as a result of the magnetorotational
instability \citep[MRI,][]{Balbus+1991} including all relativistic
effects. This is particularly important for interpreting mm-VLBI data
from the Event Horizon Telescope (EHT), which now resolves the
emission at $230$ GHz on event horizon scales. The compact size found
for Sgr A* is $\simeq 4$ Schwarzschild radii \citep{Doeleman+2008,Fish+2011}.

Total intensity images of submm synchrotron emission from such models 
predict somewhat different morphologies (size, degree of asymmetry) due to differences in the initial conditions, 
such as magnetic field configuration, electron-proton coupling, electron 
temperature distribution function and evolution, to name a few.
Although any particular model is well constrained 
\citep[e.g.,][]{Dexter+2010,Broderick+2011}, the images are often dominated by the relativistic effects 
of light bending and Doppler beaming due to an emission radius close to the event horizon, resulting 
in a characteristic crescent shape
\citep[e.g.,][]{Bromley+2001,Broderick+2006,Moscibrodzka+2009,Dexter+2010,
Kamruddin+2013,Moscibrodzka+2014,Chan+2015,Ressler+2017}. As a result, model-dependence
in parameter estimation from total intensity images is a major current issue.

The discovery of $5-10\%$ linear polarization from Sgr A* at 230 GHz \citep{Aitken+2000} showed that the
accretion rate is much less than that inferred from X-ray observations \citep{Baganoff+2001}
of hot gas at the Bondi radius \citep{Agol2000,Quataert+2000}. At the
Bondi accretion rate, the \emph{internal} Faraday rotation within the
emitting plasma should depolarize the synchrotron
radiation at 230 GHz. Later detections of \emph{external} Faraday
rotation allow an estimate of the accretion rate $\sim10^{-9}-10^{-7}
M_\odot \ \rm{yr}^{-1}$ \citep{Bower+2003,Marrone+2006}, a factor
$\simeq 100$ smaller than the Bondi value.

EHT observations provide the opportunity to measure the
\emph{spatially resolved polarization}, and show that this fraction can rise to up to $20-40\%$ on 
event horizon scales \citep{Johnson+2015}. They interpret this as
evidence for a balance of order and disorder in the underlying
polarization map, which can be well matched by maps from GRMHD
simulations \citep{Gold+2017}. 

Here we use polarized radiative transfer calculations of a single snapshot from an
axisymmetric GRMHD simulation (\S\ref{sec:models}) to understand how the resulting
polarization properties depend on the physical parameters of the
emitting plasma. We show that internal Faraday
effects become strong in a significant range of model parameter space,
scrambling and depolarizing the resulting polarization maps (\S\ref{sec:results}). Measuring the 
correlation length of the polarization direction from
spatially resolved data provides the cleanest way to set limits on the
underlying properties of the plasma. We show how this can be measured
from future EHT data as a novel constraint on the mass accretion rate
and electron temperature of the Sgr A* accretion flow.

\section{Accretion Flow and Emission models}
\label{sec:models}
 
We consider a snapshot of a 2D axisymmetric numerical solution
\citep{Dexter+2010} from the public version of the GRMHD code 
HARM \citep{Gammie+2003,Noble+2006}, where the initial conditions consist of a rotating 
black hole with dimensionless spin $a=0.9375$ surrounded by a torus in hydrostatic equilibrium \citep{Fishbone+1976} threaded 
with a weak poloidal magnetic field. The system evolves 
according to the ideal MHD equations in the Kerr spacetime.\footnote{Our snapshot is taken at time $t=2000\ GM/c^3$, where $G$ is the gravitational constant, $M$ is the mass of the black hole and $c$ is the light speed.} Turbulence due to the MRI 
produces stresses within the torus and 
leads to an outward transport of angular momentum, causing accretion of material onto the black hole.

Synchrotron radiation is produced by the hot, 
magnetised plasma and travels through the emitting medium. In the absence of any other effects, the 
resulting polarization configuration seen by a distant observer traces
the magnetic field structure of the gas.\footnote{So that the emitted polarization vector is
  perpendicular to the local magnetic field direction, we use $\rm EVPA = 1/2
  \tan^{-1}(U/Q)$, where EVPA is the electric vector position angle and Q and U are Stokes parameters.} However, as light travels the 
polarization angle is rotated both by parallel transport in the curved spacetime near the black hole and by 
Faraday rotation in the magnetised accretion flow, the latter being characterised by the Faraday rotation depth, 
$\tau_{\rho_V}=\int \rho_V dl$, where 
\begin{equation}
\rho_V=(e^3/ \pi m_{\rm e}^2c^2)\cos{\theta_B} n_{\rm{e}} B f(T_{\rm{e}} \ ,\vec{B}) / \nu^2;
\label{eq:rho_v}
\end{equation}
 $\rho_V$ is the Faraday rotation coefficient, $e, m_{\rm e}, n_{\rm{e}}$
 are the electron charge, mass and number density respectively, $\theta_B$ is
 the angle between the line of sight and the magnetic field $\vec{B}$
 with $ |\vec{B}| = B$, $c$ and $\nu$ are the light speed and
 frequency, and $f(T_{\rm{e}} \ ,\vec{B})$ is a function of $\vec{B}$
 and the electron temperature $T_{\rm{e}}$, but approximately
 $f\approx T_{\rm{e}}^{-2}$ \citep{Jones+1979,Quataert+2000}. All
 quantities are measured in the 
 comoving orthonormal fluid frame \citep{ShcherbakovHuang2011,Dexter2016}. 

MHD simulations without radiation self-consistently evolve $P_{\rm{gas}}/n
\sim T_{\rm{p}}$ and $B^2 /n$, where $P_{\rm{gas}}$ is the gas pressure, $n$ and $T_{\rm{p}}$ are the proton 
density and temperature respectively. 
Choosing the black hole mass sets the
length and timescales, while the mass accretion rate $\dot{M}$ is a free
parameter which sets the density scale. The electron temperature $T_{\rm{e}}$ 
is not self-consistently computed, and one must make a choice for it. 
Different approaches have been taken to parametrise $T_{\rm{e}}$, from a constant 
$T_{\rm{p}}/T_{\rm{e}}$ within the accretion flow \citep{Moscibrodzka+2009} to directly
evolving it with the fluid
\citep{Ressler+2015,Ressler+2017,Chael+2018} assuming some electron heating
prescription \citep{Howes2010,Rowan+2017,Werner+2018}. We assume that
$T_{\rm{e}} (\eta \ ,\alpha)= \eta  \ T_{\rm{p}} / \alpha$, with $\eta \in (0 \ ,1]$ a constant ratio between the electron and proton temperatures and $\alpha=\alpha (\mu \ ,\beta)$ a function that depends on the magnetisation of the plasma similar to the one used in \cite{Moscibrodzka+2016}: \footnote{We have taken $\mu=R_{\rm{high}}$ and $R_{\rm{low}}=1$ in the \citep{Moscibrodzka+2016} expression.}
\begin{equation}
    \alpha = \mu \frac{\beta ^2}{1+\beta ^2}+ \frac{1}{1+\beta ^2} ,
	\label{eq:Tp_Te}
\end{equation}
where the plasma parameter $\beta=P_{\rm{gas}}/P_{\rm{mag}}$ states the ratio between the gas and 
magnetic pressures and $\mu$ is a free parameter that describes the electron to proton coupling in 
the weakly magnetised zones (disc body) of the simulation. 

The numerical solution we use has a Blandford-Znajek jet \citep[]{McKinney2006}. 
Different choices of $\mu$, the electron-proton coupling factor in Eq. \ref{eq:Tp_Te}, can cause the wall between the accretion flow and the jet to shine. 
When $\mu$ in equation \ref{eq:Tp_Te} is small, the disc has a very high temperature and lights up at 230 GHz 
due to the fact that, compared to the jet, it has both the highest density and magnetic field strength. However, the 
larger the $\mu$ the colder the disc is and the fainter it gets. If one wishes to maintain a fixed flux, the accretion rate onto the black hole $\dot{M}$ must increase. As a consequence, the jet wall can light up first even given its lower density and field strength. 

For a given choice of  $\eta$ and $\mu=(1 \ ,2 \ ,5 \ ,10 \ ,40 \
,100)$, $\dot M$ is then chosen in such a way that the total flux
$F_\nu$ at 230 GHz is either $3 $ Jy or $0.3$ Jy. We chose the first
value to model Sgr A* and the second, $0.3$ Jy, arbitrarily to
decrease $\alpha_I/\rho_V$, where $\alpha_I$ is the total
absorption coefficient. This second option for $F_\nu$ gives us the
opportunity to study models in the optically thin regime to separate
the effects of absorption and Faraday rotation.

Given that $n_{\rm{e}} \propto \dot{M}$, $B \propto \dot{M}^{1/2}$,
$F_\nu \propto n_{\rm{e}}^\xi B^\kappa {T_{\rm{e}}}^\sigma$ (where
typically $\xi \in [0,1]$, $\kappa \in [0,2]$, $\sigma \in [1,4]$),
and assuming a constant $F_\nu$, we can express the Faraday rotation
depth, $\tau_{\rho_V}$, as a function of $\dot{M}$:
\begin{equation}
\tau_{\rho_V} \sim n_{\rm{e}} B {T_{\rm{e}}}^{-2}  \propto \dot{M}^{\delta};
\label{eq:tau_rho_v}
\end{equation}
%
where $\delta \equiv 3/2 + (2\xi + \kappa)/ \sigma \simeq 3/2-7/2$. It
can be seen from eq. \ref{eq:tau_rho_v} that $\tau_{\rho_V}$ has a
strong dependence on $\dot{M}$ and small changes of this quantity
reflect as big differences in $\tau_{\rho_V}$. Due to this, models
with similar physical parameters can vary widely in the strength of
the internal Faraday effects and, for $\tau_{\rho_V} \gtrsim 1$, in the
resulting polarization structure.

To account for emission, absorption, parallel transport and Faraday effects locally within the accretion 
flow, we employ the publicly available numerical code \textsc{grtrans}  \citep{Dexter+2009,Dexter2016}\footnote{http://www.github.com/jadexter/grtrans} to do a self-consistent 
fully relativistic ray tracing radiative transfer calculation at $230$ GHz. 
The output of the calculation is a polarized image of our GRMHD
snapshot as seen by a distant observer at a 50 degree inclination from
the black hole (and accretion flow) rotation axis.

\section{Results}
\label{sec:results}
\begin{figure*}
	\includegraphics[trim = 0cm 0cm 2.5cm 2cm, clip=true,width=\columnwidth]{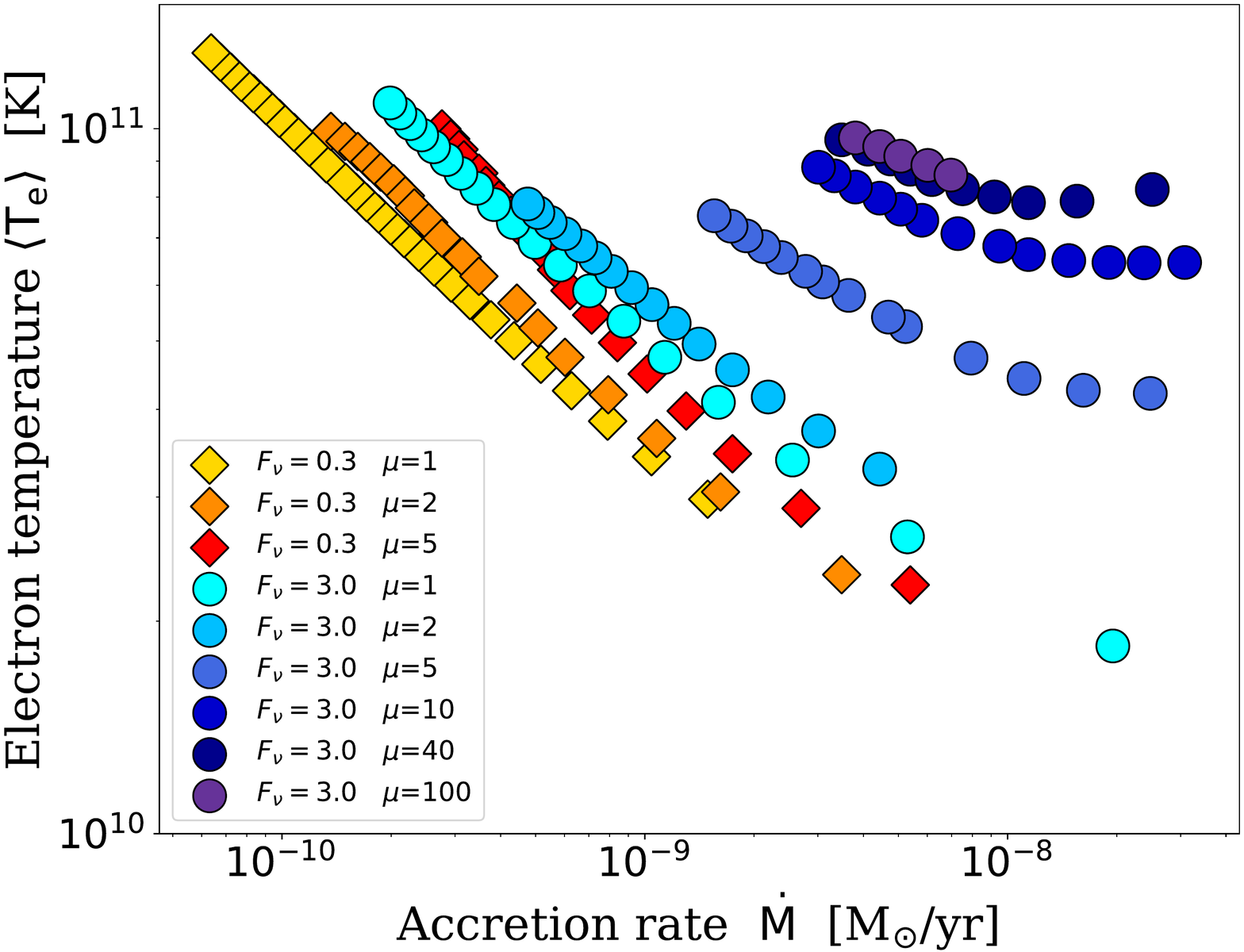}
	\includegraphics[trim = 0cm 0cm 2.5cm 2cm, clip=true,width=\columnwidth]{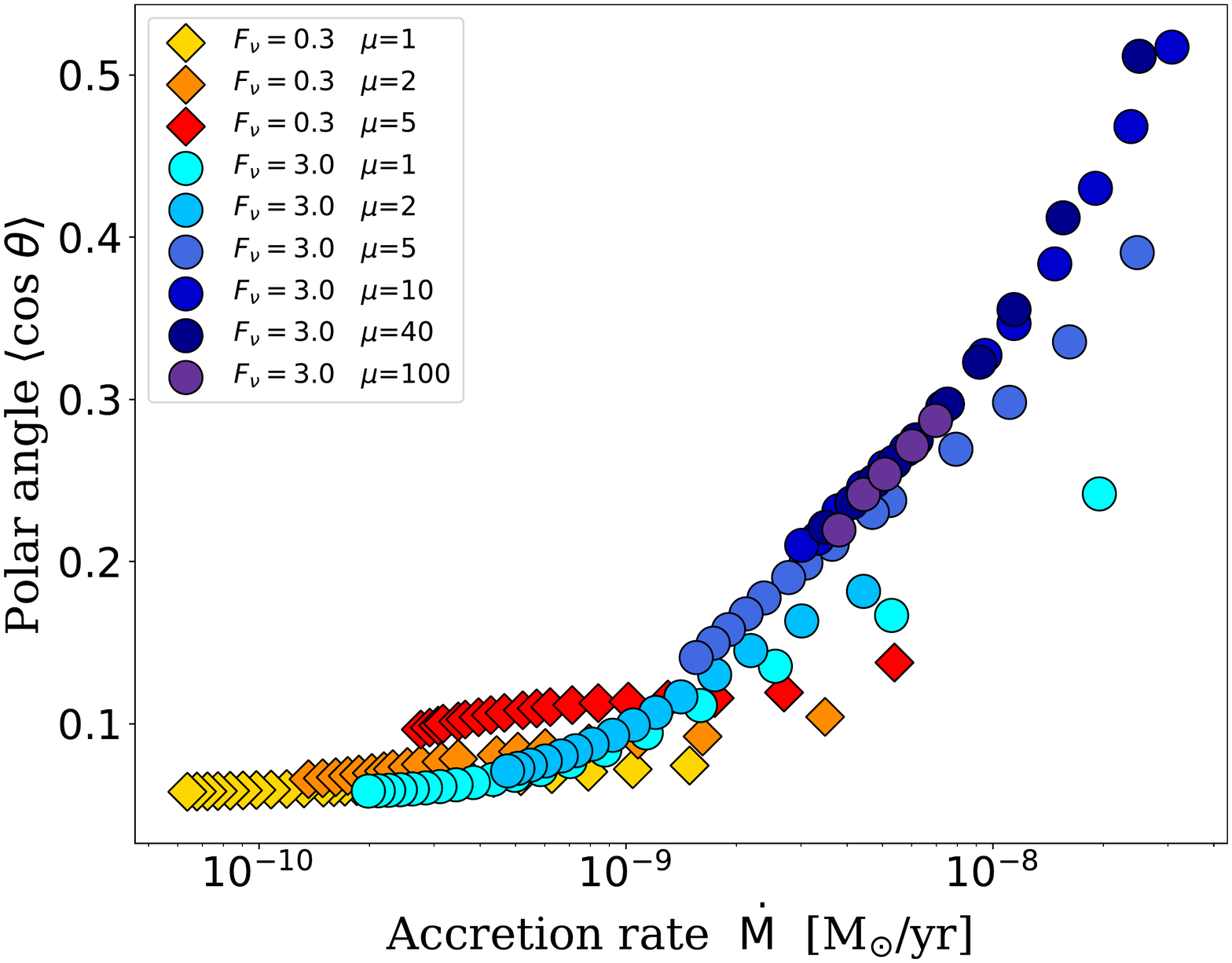}
	\includegraphics[trim = 0cm 0cm 2.5cm 1.5cm, clip=true,width=\columnwidth]{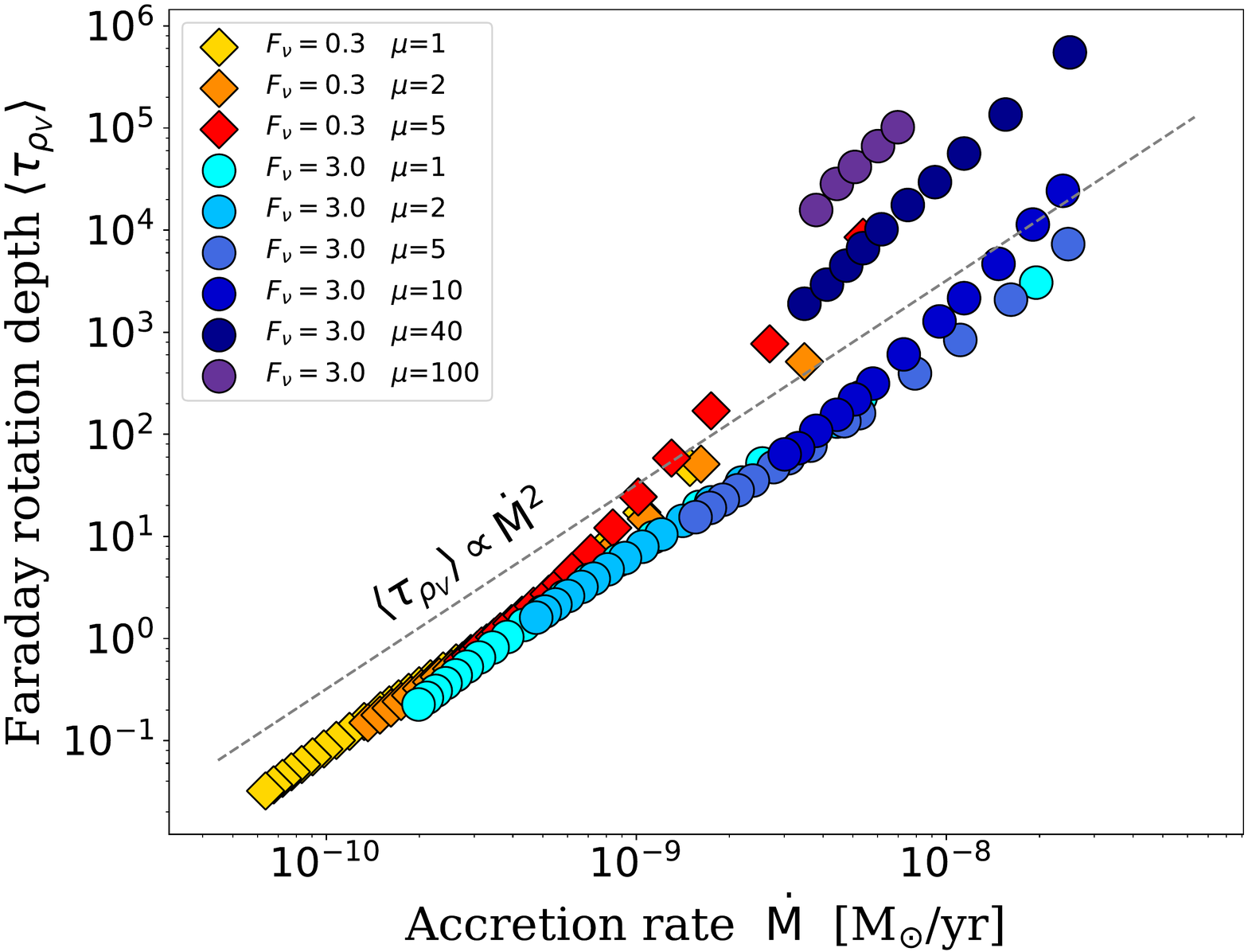}
    \caption{Intensity-weighted, image-integrated model quantities. Each dot is a chosen ($\dot{M},T_{\rm{e}}$) pair with $F_\nu$ indicated by a colour scale: diamonds for $F_\nu=0.3$ Jy and circles for $F_\nu=3$ Jy. The colour gradient shows the choice for $\mu$, where the lighter (darker) the shade, the lower (higher) the $\mu$ value is. \textbf{Upper left}:  Intensity-weighted, image-averaged electron temperature $\langle T_{\rm{e}}\rangle$ \emph{vs} $\dot{M}$. The models that show a steady decrease of $\langle T_{\rm {e}} \rangle$ with $\dot{M}$ are more ``disc-like'' systems. The transition to ``jet-like'' systems happens where the decrement stops and becomes constant with $\dot{M}$. \textbf{Upper right}: Intensity-weighted image-averaged emission angle $\langle \cos \theta \rangle$ as a function of $\dot{M}$. At high $\dot{M}$, as $\mu$ increases the emission region moves towards the poles, indicating the transition to a more ``jet-like'' system. \textbf{Bottom}: Intensity-weighted image-averaged Faraday rotation depth $\langle \tau_{\rho_V}\rangle$ \emph{vs} $\dot{M}$. It can be seen how $\langle \tau_{\rho_V}\rangle$ extends over a wide range of values with $\dot{M}$. A scaling relation between both quantities is shown with a dotted line where $\delta=2$.}
    \label{fig:te_mdot}
\end{figure*}
Fig. \ref{fig:te_mdot} shows the resulting intensity-weighted, 
image-averaged electron temperature $\langle T_{\rm {e}} \rangle$, polar angle $\langle \cos
\theta \rangle$, and Faraday rotation depth $\langle \tau_{\rho_V} \rangle$ for each
input model with varying ($\dot{M},T_{\rm{e}}$). 
The steady decrease
of $\langle T_{\rm {e}} \rangle$ with $\dot{M}$ in the upper left panel of
Fig \ref{fig:te_mdot} points to ``disc-like'' systems. The transition
to ``jet-like'' systems happens when the circles show a constant
behaviour with $\dot{M}$ and is highly dependent on $\mu$. As
discussed before, at large $\mu$ the jet has a high enough temperature
that it can outshine the cold disc.

This is shown as well in the right panel of Fig. \ref{fig:te_mdot}, where the cosine of the inclination angle where most of the emission comes from, $\langle \cos \theta \rangle$, as a function of $\dot{M}$ is plotted. It can be seen that at high $\dot{M}$, models that have the same accretion rate but different $\mu$ values have different emission regions. As the electron-proton coupling $\mu$ increases, the emission region moves towards the poles indicating a transition to a more ``jet-like'' system.

In the bottom panel of Fig. \ref{fig:te_mdot} we plot the intensity-weighted image-averaged 
Faraday rotation depth $\langle \tau_{\rho_V} \rangle$ \emph{vs} $\dot{M}$ and the scaling
relation in eq. \ref{eq:tau_rho_v}, with $\delta=2$.\footnote{We included the effects of Faraday conversion in the radiative transfer calculation as well, but it is significantly weaker than Faraday rotation, and so we focus our study to the latter.} 
This panel shows very nicely the wide spread in $\langle \tau_{\rho_V} \rangle$ values as a
function of $\dot{M}$, supporting the idea that systems with similar
physical parameters can have widely varying strengths of Faraday
rotation. This makes $\langle \tau_{\rho_V} \rangle$ a sensitive tracer of the
physical conditions of the plasma. 

\begin{figure*}
	\includegraphics[trim = 5cm 0cm 6.5cm 2.1cm, clip=true,width=\columnwidth]{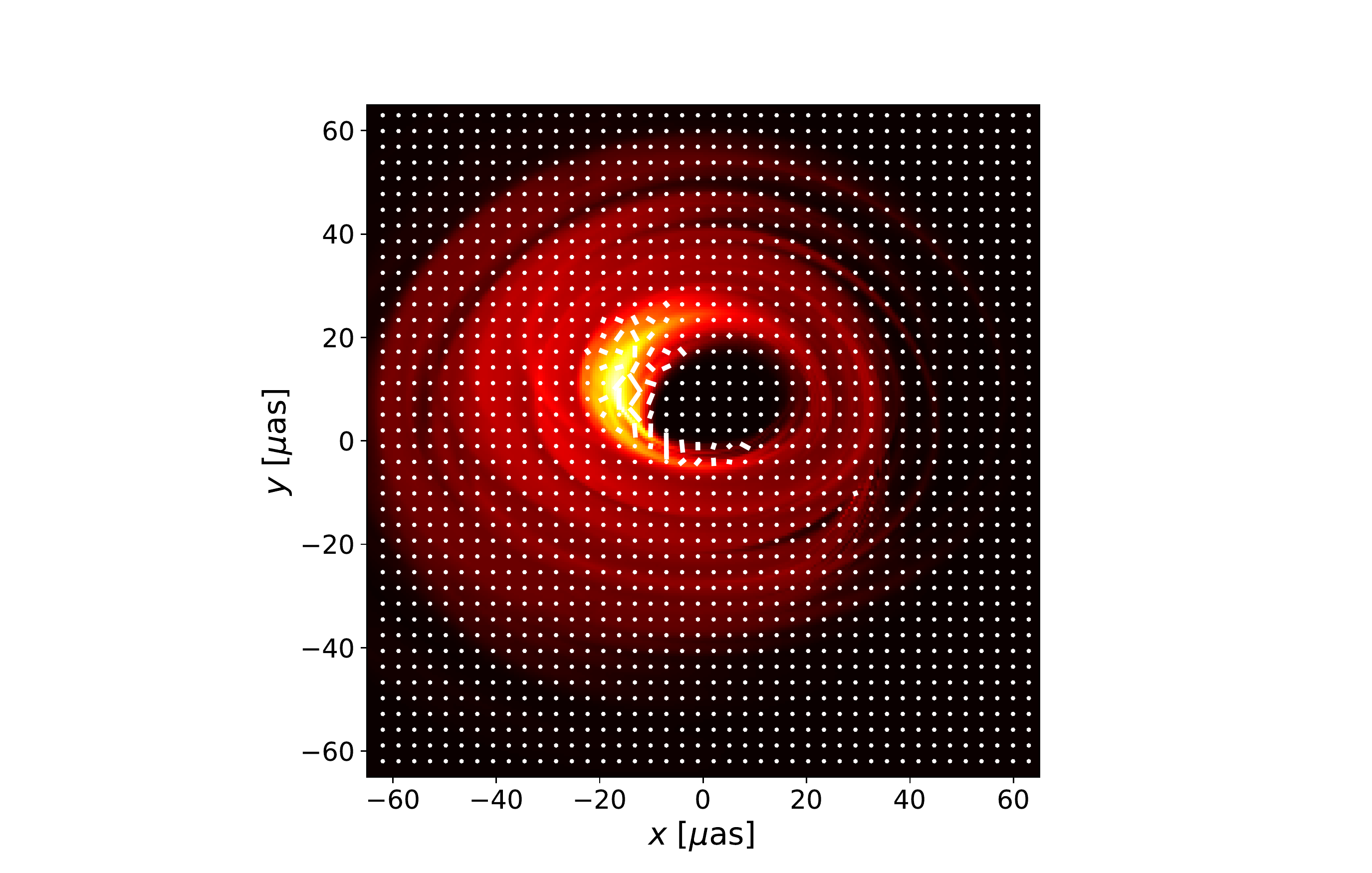}
	\includegraphics[trim = 5cm 0cm 6.5cm 2.1cm, clip=true,width=\columnwidth]{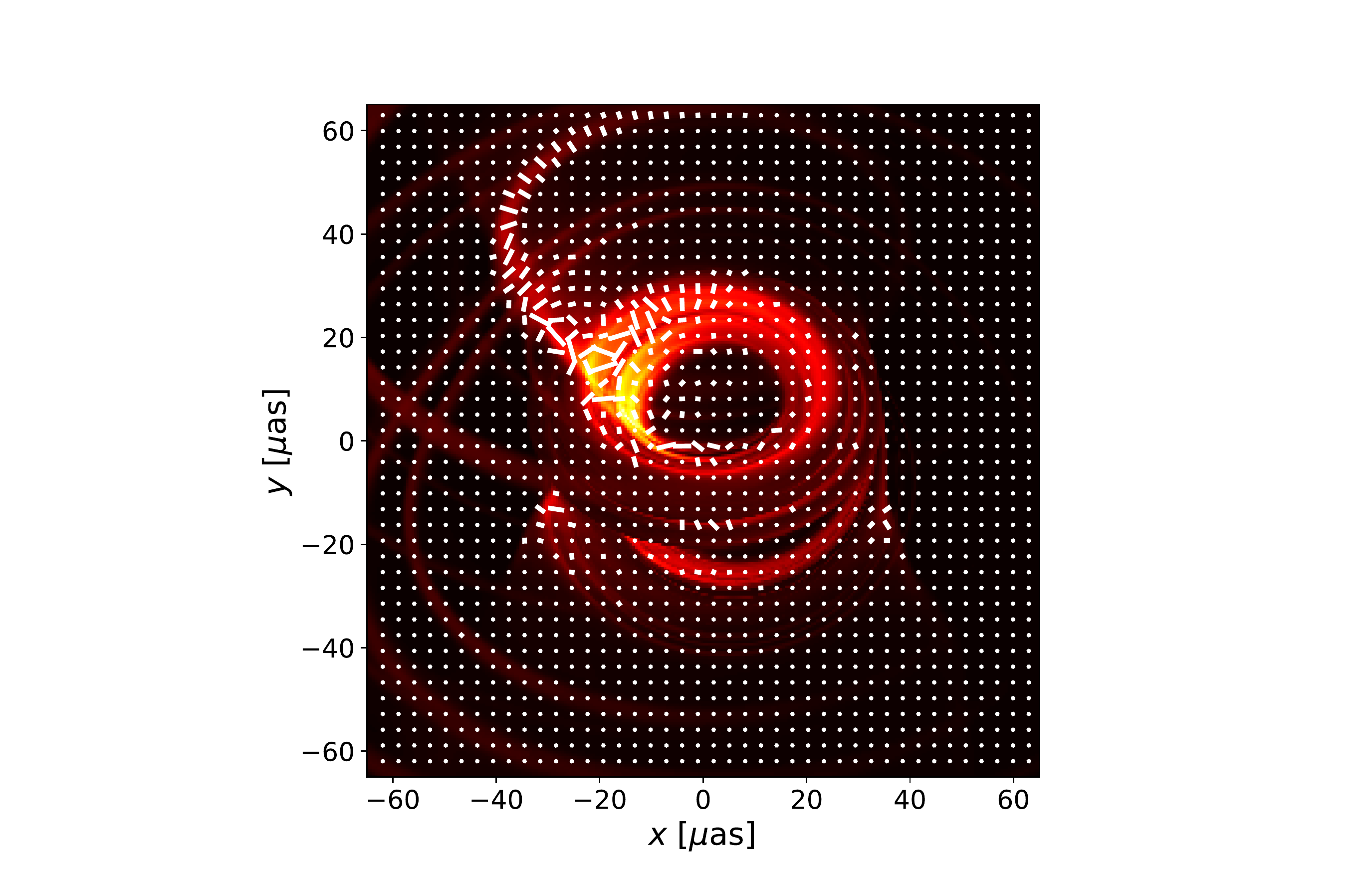}
	\includegraphics[trim = 5cm 0cm 6.5cm 2.1cm, clip=true,width=\columnwidth]{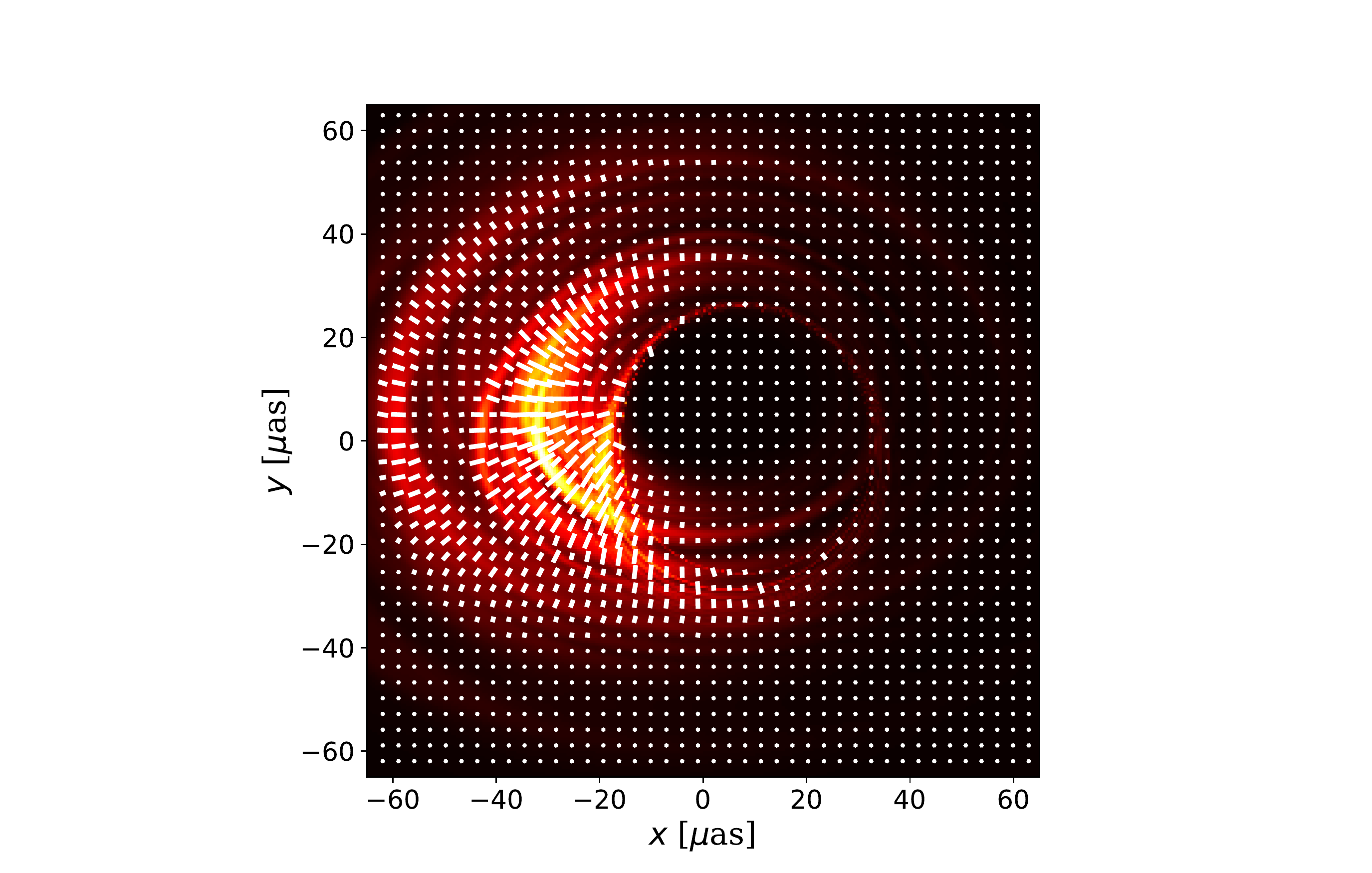}
	\includegraphics[trim = 5cm 0cm 6.5cm 2.1cm, clip=true,width=\columnwidth]{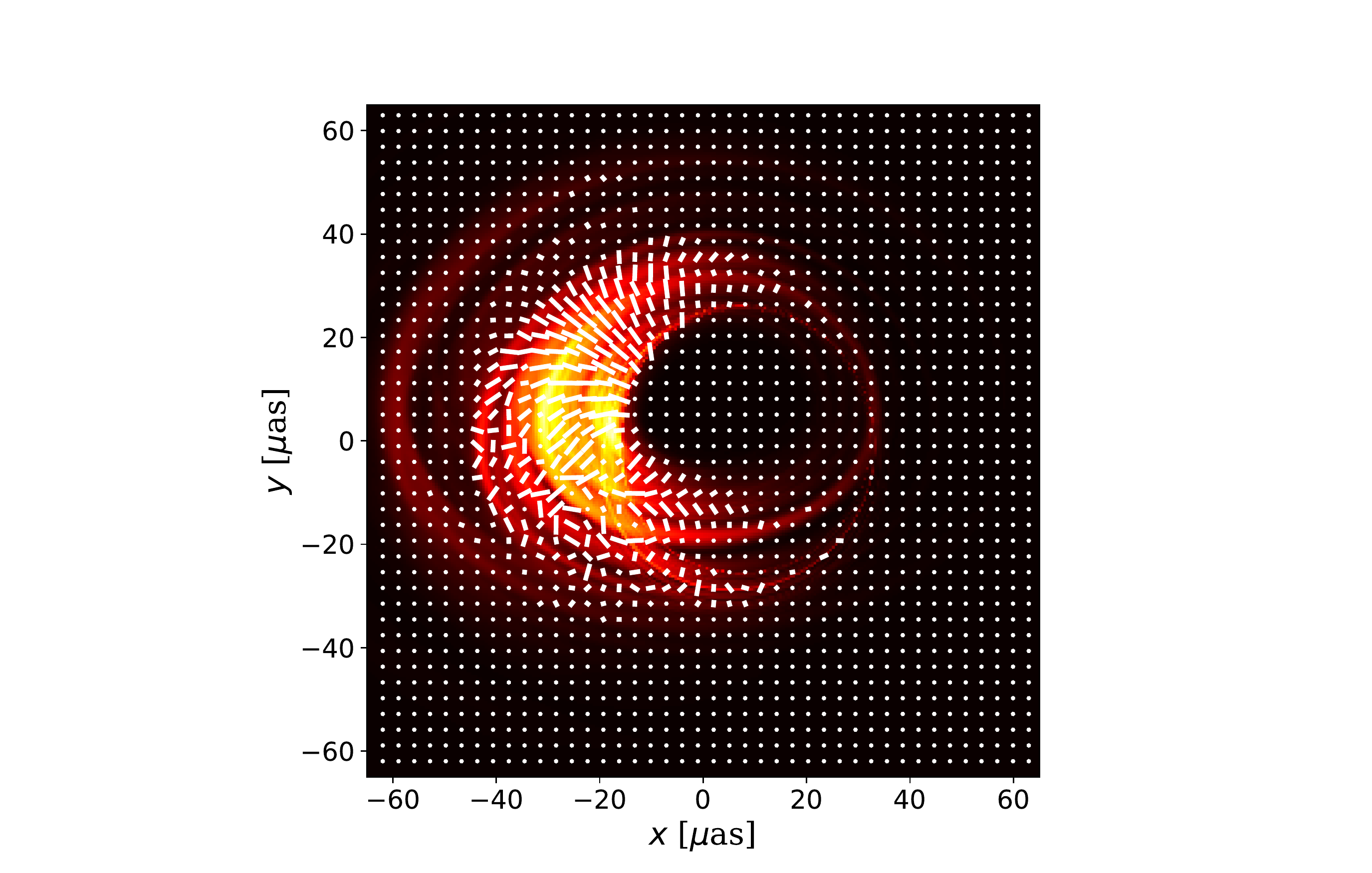}
\caption{ Polarization maps obtained from different $\mu$ and
  ($\dot{M},T_{\rm{e}}$) pairs. It can be seen how the Faraday effects
  affect the polarization. \textbf{Upper left panel}: optically thick
  image. The accretion flow emits like a blackbody and is
  depolarized. \textbf{Upper right panel}: depending on the choice for
  the electron to proton coupling, the wall between the
  Blandford-Znajek jet and the accretion flow may become
  apparent. \textbf{Bottom left panel}: weak Faraday effects. The ticks
  trace the smooth magnetic field configuration. \textbf{Bottom right panel}: strong Faraday effects. The ticks are disordered and the underlying magnetic field structure is less evident. }
\label{fig:PM}
\end{figure*}

Fig. \ref{fig:PM} shows four sample polarized images. Plotted in the background is the total 
intensity image centred on the black hole (colour shows total flux on a linear scale where the lighter the shade 
the greater the emission).

These images show the accreting material in characteristic asymmetric crescent shapes: due to Doppler 
beaming in the rotation torus, the left side where the gas approaches the observer is much brighter than the 
right side. Furthermore, strong relativistic light bending lets us see behind the black hole, whose emission appears 
to be coming from above and below it in the images. In the foreground, white ticks show linear polarization fraction (LP)
direction with their length proportional to the LP magnitude (given by $\sqrt{Q^2+U^2}$). This is what we refer to as a polarization map (PM). 

The images shown in Fig. \ref{fig:PM} have a variety of model parameters. In the upper left panel an optically thick image is shown. In this case, the system resembles a black body, dominated by optical absorption and is completely depolarized from self-absorption. We can see a case of the Blandford-Znajek jet wall lighting up in the upper right panel of Fig. \ref{fig:PM} for a case where $\mu=100$.

The bottom panels of Fig \ref{fig:PM} are optically thin.
In the case of weak Faraday effects (hot, tenuous emitting medium -
high $T_{\rm{e}}$ and low $\dot{M}$, $\langle \tau_{\rho_V} \rangle<1$; bottom left
panel), the polarization traces the toroidal magnetic field
(horizontal where the light comes from the approaching gas and
vertical where it comes from gas behind the black hole) and displays
an ordered behaviour (due to the axisymmetry of the system). The
combination of this with the crescent shape background image leads to
such a characteristic polarization map \citep{Bromley+2001}. When
considering the polarization over the whole image, the contributions
from each of the vector components may cancel, resulting in a lower LP
over the image (\emph{beam depolarization}). Strong Faraday effects
(colder, denser medium - low $T_{\rm{e}}$, high $\dot{M}$; bottom
right panel of Fig. \ref{fig:PM}) can scramble and depolarize the
image on small scales.

\subsection{Linear Polarization Fraction and Rotation Measure in the models.}
\begin{figure*}
	\includegraphics[trim = 0cm 0cm 2.3cm 0.5cm, clip=true,width=\columnwidth]{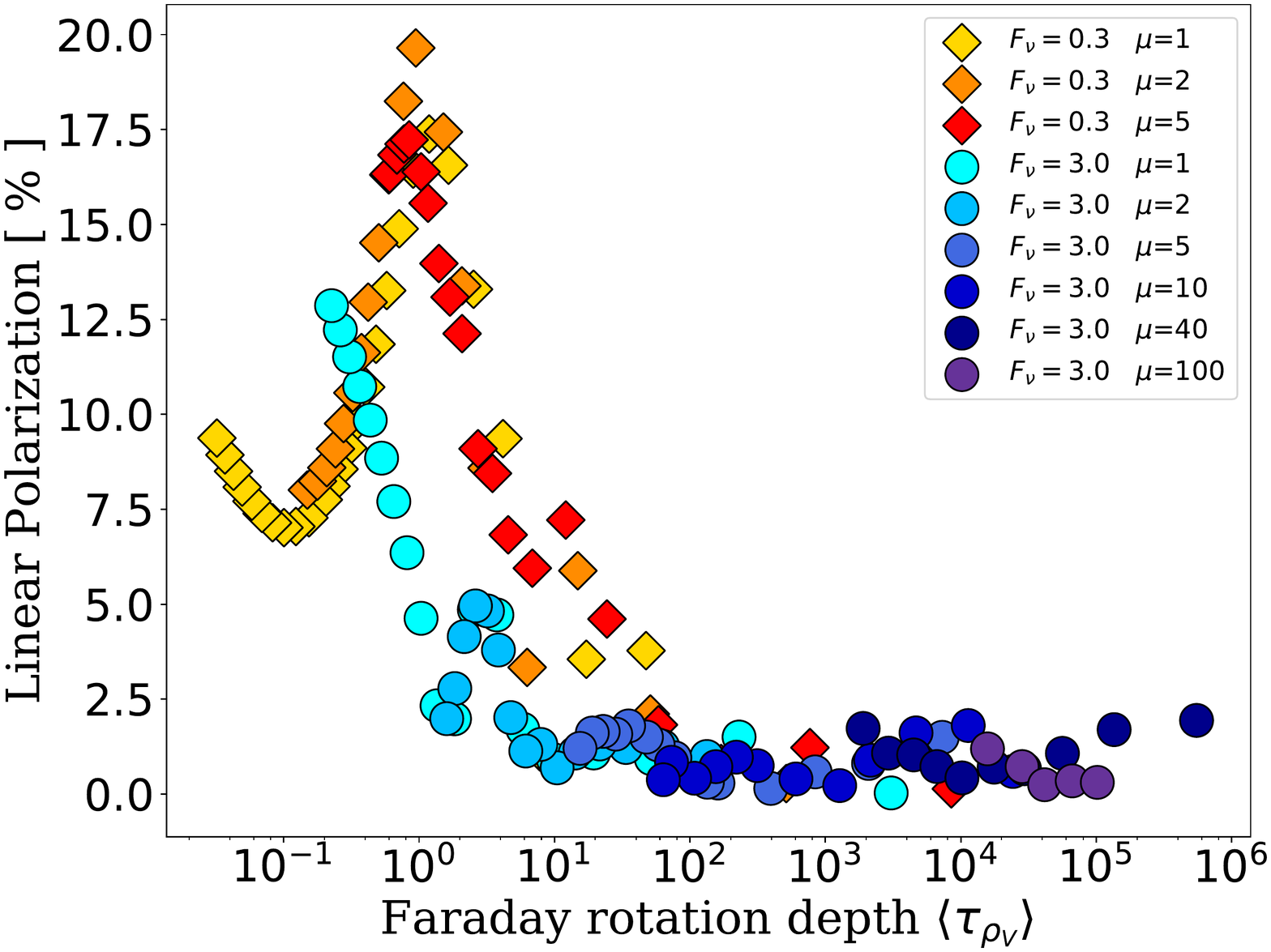}
    \caption{Net linear polarization fraction plotted against the intensity-weighted image-averaged 
      Faraday rotation depth $\langle \tau_{\rho_V} \rangle$. The same colour and
      marker criteria has been used as that in figure
      Fig. \ref{fig:te_mdot}. As the Faraday effects become stronger,
      the LP decreases, as expected. However the behaviour is neither
      smooth nor universal ( $\langle \tau_{\rho_V} \rangle \lesssim
      10^2$). ``Jet-like'' models have high Faraday optical depths
      ( $\langle \tau_{\rho_V} \rangle \gtrsim 10^2$) from the cold, dense disc and are heavily
      depolarized, failing to reproduce the Sgr A* LP of $\simeq 5-10\%$.}
    \label{fig:lin_pol_frac}
\end{figure*}

Fig. \ref{fig:lin_pol_frac} shows the net intensity-weighted LP integrated over the image
versus the intensity-weighted image-averaged Faraday rotation depth, $\langle \tau_{\rho_V} \rangle$. The images generally depolarize with increasing $\langle \tau_{\rho_V} \rangle$, as expected. However, the individual behaviour between both sets of models (diamonds and circles) is different, and no smooth or uniform LP trend as a function of $\langle \tau_{\rho_V} \rangle$ can be extracted. 

Given a measurement of LP, one could use Fig. \ref{fig:lin_pol_frac}
to set an upper limit on $\langle \tau_{\rho_V} \rangle$ and obtain the models that
satisfy the restriction set by the LP. As an example, a variety of our
``disc-like'' models where $\langle \tau_{\rho_V} \rangle$ varies over many orders of
magnitude ($\sim5\times10^{-2}-1\times10^2$) satisfy the measured
$5-10\%$ LP for Sgr A* at 230 GHz \citep{Aitken+2000}. The emission is
locally strongly polarized ($\gtrsim 40\%$) but is naturally beam
depolarized due to the combination of the crescent image and toroidal
magnetic field configuration. The ``jet-like'' models ( $\langle \tau_{\rho_V} \rangle
\gtrsim 10^2$) on the
other hand are heavily depolarized, and cannot match the observed LP
of Sgr A*.

We looked at the dependence of the rotation measure (RM) of the images with 
$\langle \tau_{\rho_V} \rangle$ as well and found, as expected, that the RM 
increases with $\langle \tau_{\rho_V} \rangle$. However, we could not get a clean
measurement like in \cite{Bower+2003} and \cite{Marrone+2006} because
our simulation domain is not as large as that in their work. We can
look at this in the future, but it would require very long duration
simulations \citep[e.g.,][]{Narayan+2012} to reach inflow equilibrium
at the large radii of the external Faraday screen.

\subsection{The Correlation Length.}
\begin{figure*}
\includegraphics[trim = 2cm 0cm 3.2cm 0cm, clip=true,width=\linewidth]{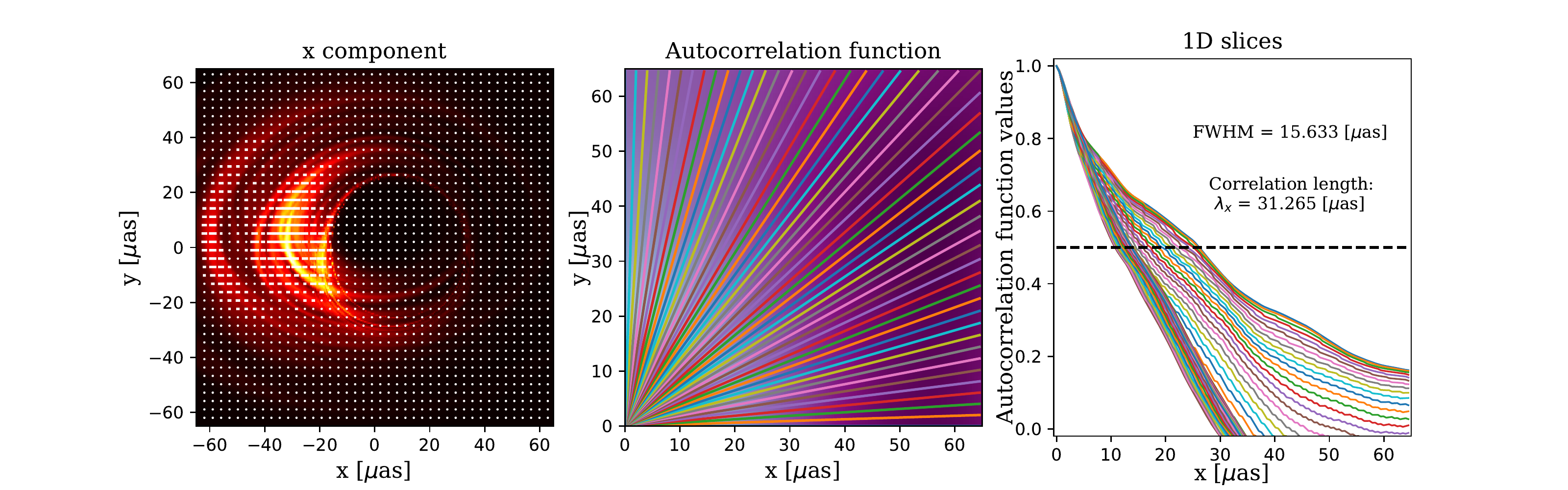}
\caption{Illustration of the calculation of the polarized correlation
  length. \textbf{Left:} take one of the vector components of the
  polarization ($x$ component shown here) and auto-correlate the map. \textbf{Middle:} Plotted in the background in shades of purple is the 2D autocorrelation function. We take 1D slices of this function in different angular directions (coloured solid lines in the foreground) to account for the 2D behaviour. \textbf{Right:} 1D slices from the autocorrelation function. Twice the average of their values at 0.5 (FWHM) is how we define as the polarized correlation length, $\lambda_{x}$, in $\mu$as.}
\label{fig:corr_length_method}
\end{figure*}

The upper panels of Fig \ref{fig:PM} can be easily distinguished from their total intensity images alone 
and might be disfavoured already from the measured size of the source and spectral observations. 
In the optically thin regimes however (bottom panels), the total intensity images are hard to tell apart and 
are generally consistent with the observational constraints \citep{Moscibrodzka+2009,Dexter+2010}. The 
polarization maps however, vary substantially. This spatial configuration of the polarization offers an alternative 
to learning about the physical parameters of the models. 

To characterise the degree of order of each map produced by a particular ($\dot{M}$,\ $T_{\rm{e}}$) pair, we use a quantity which we call the ``correlation length'', $\lambda$.
Large values of $\lambda$ point to an ordered configuration, limited by the coherence of the magnetic field 
structure, whereas small values indicate a more disordered configuration.

To calculate this quantity we autocorrelate each map. Because the PM is a vector field, we look 
at each component separately and weight their value at each pixel by Stokes $I$ at the same pixel\footnote{The 
reference system is orthonormal with one of the axes aligned with the spin axis of the black hole and the other in a direction perpendicular to the observer \citep[]{Bardeen+1972}.} (left panel of Fig. \ref{fig:corr_length_method}). The result is a 2D function 
that gives information on how the polarization component varies spatially. We then take 50 ``1D slices'' of this 
function in different angular directions to account for the spatial changes in 2D and take the average of their widths at 0.5 (FWHM, middle and right panels of Fig. \ref{fig:corr_length_method}). Twice this value is the polarized correlation length in $\mu$as.

\begin{figure*}
	\includegraphics[trim = 0cm 0cm 2.3cm 0.5cm, clip=true,width=\columnwidth]{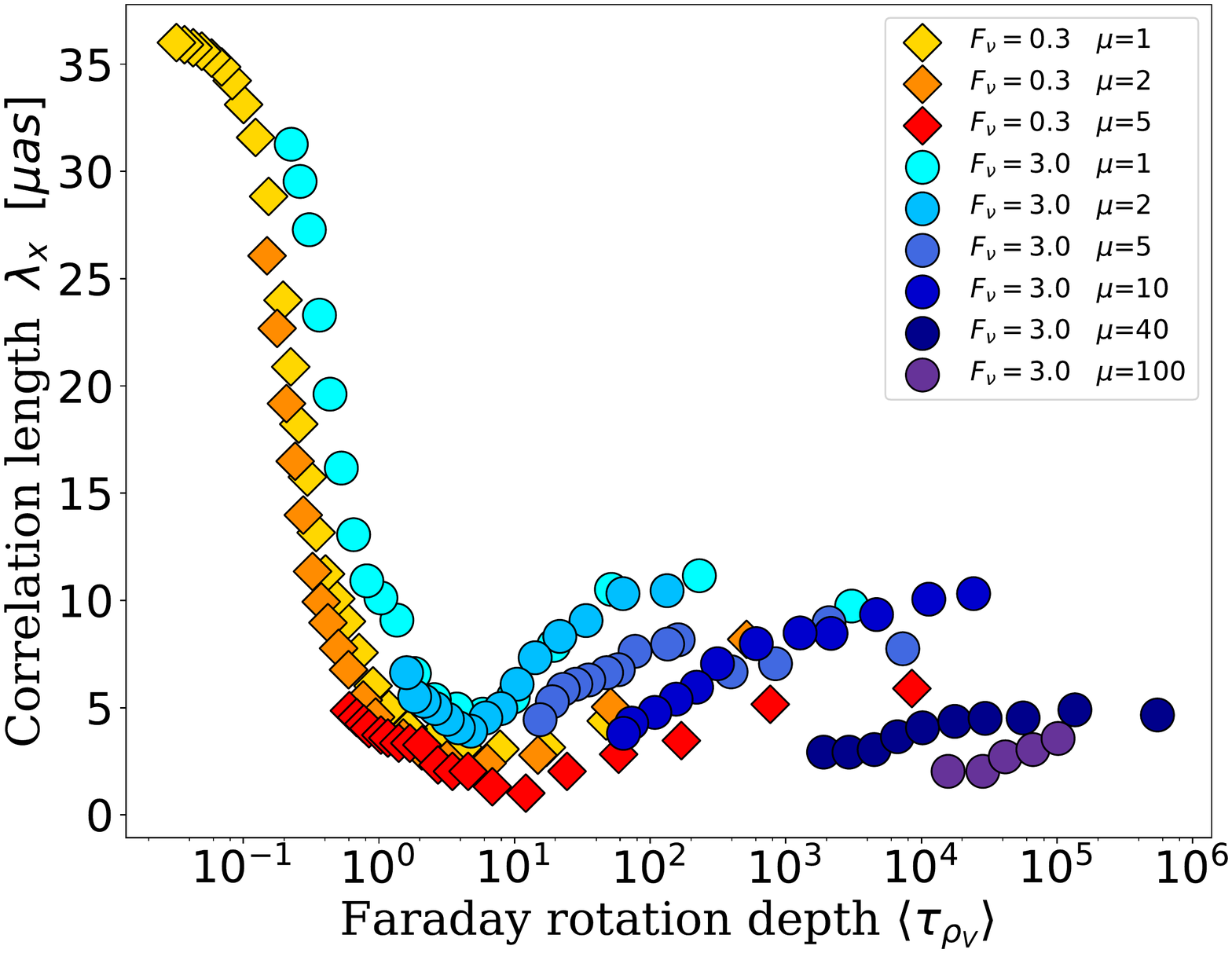}
	\includegraphics[trim = 0cm 0cm 2.3cm 0.5cm, clip=true,width=\columnwidth]{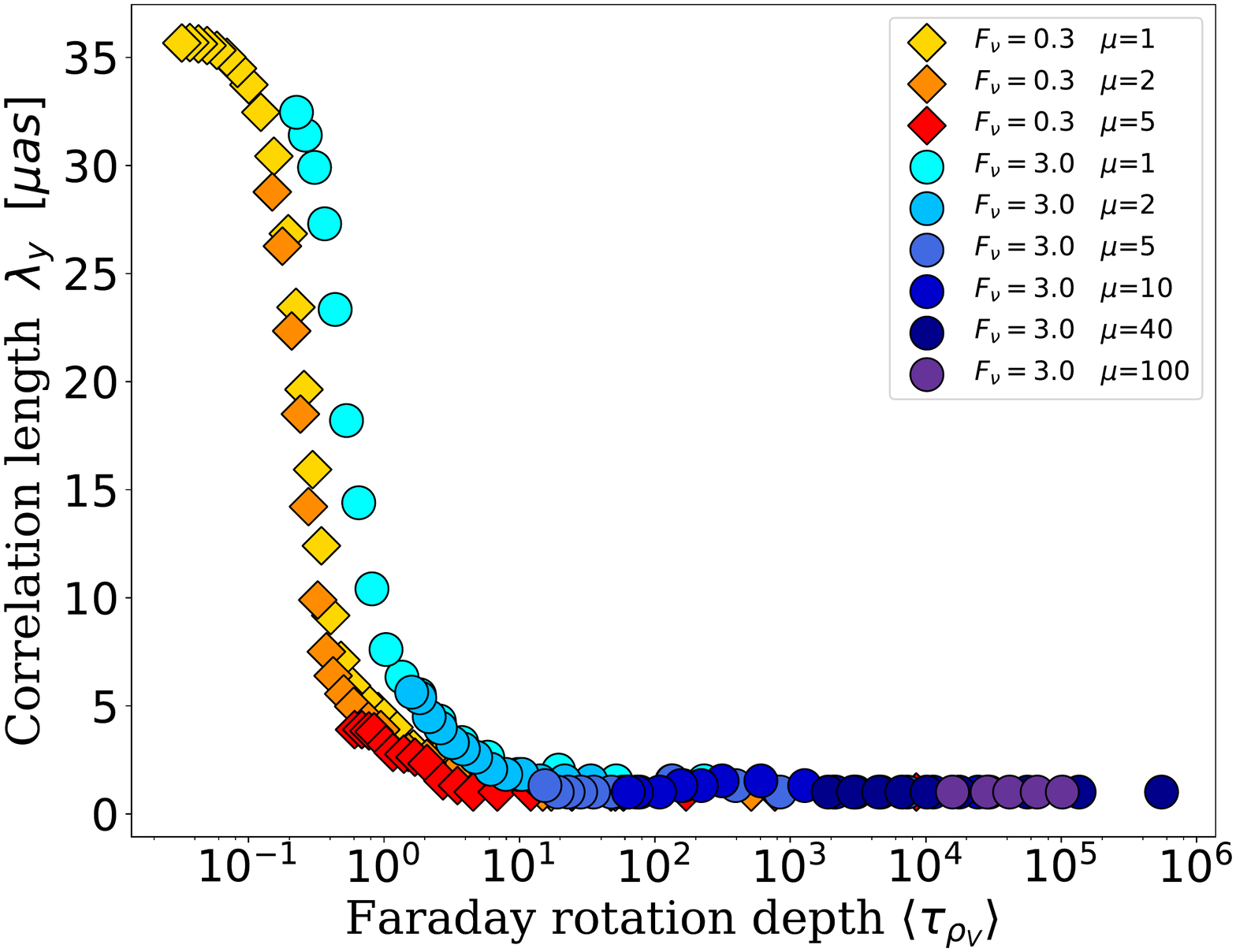}
    \caption{Correlation length measured for $x$ and $y$ vector components (left and right panels respectively) of 
    PMs with different $\mu$ and ($\dot{M}$,$T_{\rm{e}}$) values
    plotted against the intensity-weighted image-averaged Faraday rotation depth
    $\langle \tau_{\rho_V} \rangle$. The same colour and marker criteria is used as
    that in Fig. \ref{fig:te_mdot} and
    Fig. \ref{fig:lin_pol_frac}. Coherent maps are obtained when
    $\langle \tau_{\rho_V}\rangle \lesssim1$ and scrambling appears as the Faraday
    effects become stronger. A measurement of the correlation length
    places a model-independent upper limit on $\langle \tau_{\rho_V} \rangle$, and in
    turn the lower limits on the plasma electron temperature and
    relative magnetic field strength.}
    \label{fig:corr_length}
\end{figure*}

Figure \ref{fig:corr_length} shows the correlation length ($\lambda_x$
and $\lambda_y$, $x$ and $y$ subindices for each vector component) of
each simulation's PM as a function of the intensity-weighted image-averaged Faraday rotation depth
$\langle \tau_{\rho_V} \rangle$. 

The overall behaviour of the correlation length for both components is as expected. At small $\langle \tau_{\rho_V} \rangle$ 
(small $\dot{M}$, large $T_{\rm{e}}$), the Faraday effects are weak implying coherent PMs in which the changes 
are given by the geometry of the magnetic field in the gas, resulting in a maximum of the correlation length. 
As  $\langle \tau_{\rho_V} \rangle$ increases, the Faraday effects become stronger, the scrambling becomes more apparent 
and the correlation length decreases, showing a sharp drop at around $\langle \tau_{\rho_V} \rangle \approx 1$.
In the case of the correlation length for the $x$ component, there is a small increase after the minimum that remains 
until high values for $\langle \tau_{\rho_V} \rangle$ which we associate with the internal structure of the polarization. This does not appear in the $\lambda_y$ plot.

Unlike the LP (Fig. \ref{fig:lin_pol_frac}), the $0.3$ Jy and $3$ Jy correlation length curves have the same shape and eventually overlap, which points towards a universal behaviour for this quantity (in the limited range of models studied so far). This has powerful implications, since
measuring $\lambda$ sets a model-independent upper limit on $\langle \tau_{\rho_V} \rangle$($\dot{M}, T_{\rm{e}},\beta,\rm{H/R})$, where $\rm{H/R}$ is the scale height of the disc.

One can relate the correlation length directly to observable quantities in VLBI \citep{Johnson+2015}.
In their paper, \citet{Johnson+2015} show examples with Gaussian intensity 
distributions and constant net LP (~$6-7\%$), and a varying polarization structure with a prescribed coherence length. They find a correlation length of 0.29 times the Gaussian FWHM of the model. 
Measuring an approximate Gaussian FWHM for our images and multiplying it by 0.29 gives us an estimated correlation length of ~$11.6\mu$as. We then use Fig. \ref{fig:corr_length} to set an upper limit on the $\langle \tau_{\rho_V}\rangle \lesssim1$, in agreement with the qualitative argument of \citet{Agol2000} and \cite{Quataert+2000}. 
However, our models are not Gaussians and it is not clear that the
$\lambda$ value inferred by \cite{Johnson+2015} applies here.

We can extend the analysis further into visibility space.
\begin{figure*}
	\includegraphics[trim = 2.5cm 2cm 3cm 1cm, clip=true,width=\linewidth]{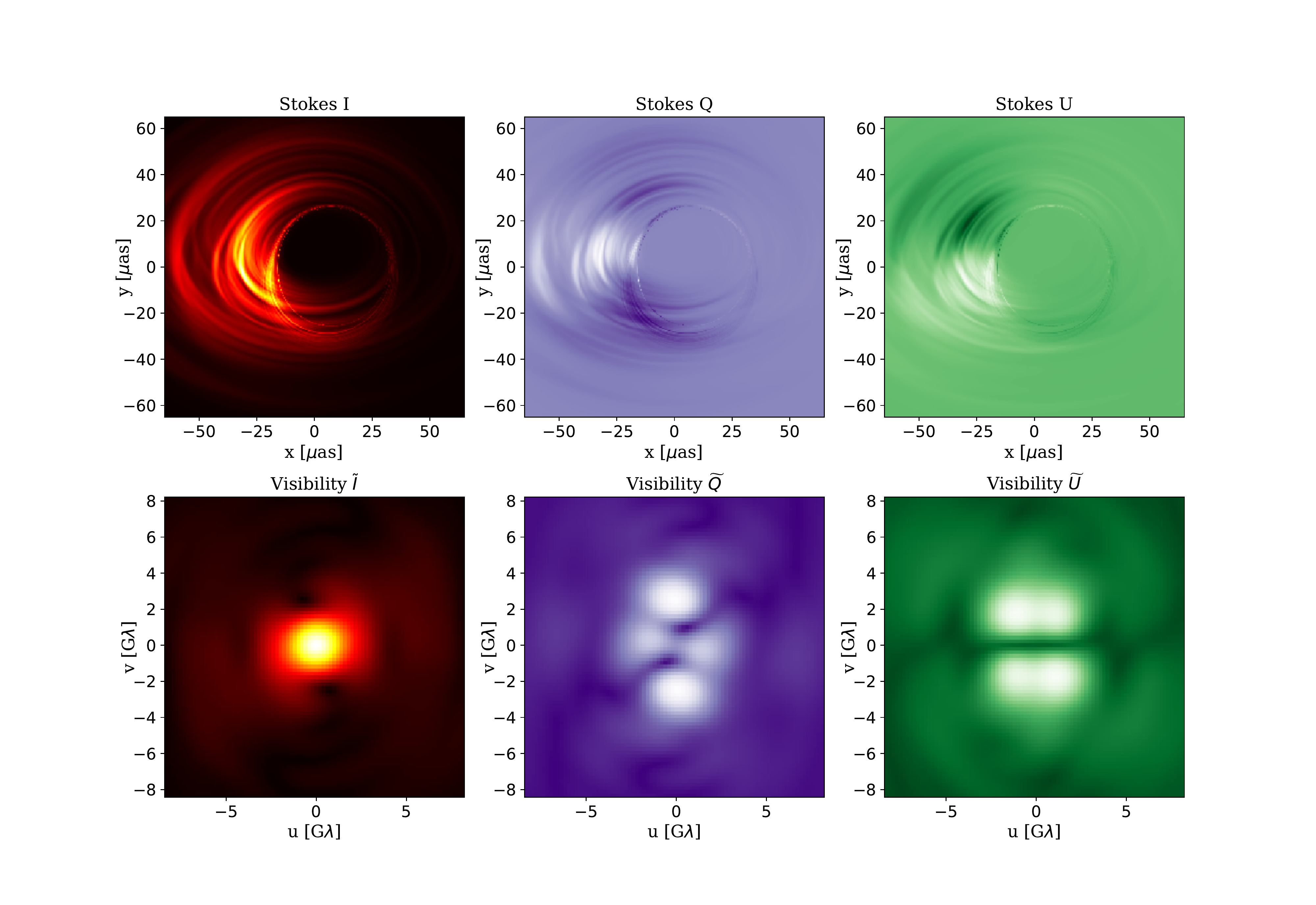}
    \caption{$I$, $Q$ and $U$ Stokes parameters for one of our
      simulations in image (top panels) and visibility space (bottom
      panels). On the top images, it can be seen that $Q$ and $U$
      resemble $I$ on large scales, with different substructure due to
      the changing polarization. On the visibility space however
      (bottom panels), small scale features in $\widetilde{Q}$ and
      $\widetilde{U}$ give information on $I$ whereas the large scale
      structure corresponds to the polarization.}
    \label{fig:Q_image}
\end{figure*}
Fig. \ref{fig:Q_image} shows the Stokes parameters $I$, $Q$ and $U$ (top) and their respective visibilities $\widetilde{I}$, $\widetilde{Q}$ and $\widetilde{U}$ (bottom) of one of our simulations.
On large scales, it can be seen that $Q$ and $U$ resemble $I$, showing the same crescent structure. However, on smaller scales some deviation becomes evident because the polarization, $\vec{p}$, is changing. Therefore, on the largest scales one gets information on $I$ and on the smallest scales one sees the polarization properties.

If we study this in the visibility space $uv$ and take the Fourier Transform ($FT$) of the Stokes parameters (bottom images of Fig. \ref{fig:Q_image}), the roles are inverted. Large scale features become small and vice versa. 
In this respect, the shape of the total intensity image becomes a small ``beam'' in $uv$, whereas the large scale structure observed in the $\widetilde{Q}$ and $\widetilde{U}$ images corresponds to the smallest angular scales in $Q$ and $U$ and reflects the properties of the polarization map.

One can think of the images as the convolution of $I$ with $\vec{p}$, with the result interpreted as $I$ being smeared out by $\vec{p}$ with some characteristic scale that reflects the inner structure of the latter. 
In the case of a completely disordered polarization map, taking the $FT$ of the image would give what would basically be a noise map in the visibility space, with no characteristic scale at which the polarization's behaviour stands out.
On the other hand, the $FT$ of the convolution between $I$ and a completely ordered $\vec{p}$ would give an image with a nice beam centred at $uv=0$ and no noise whatsoever. 

We are interested in finding the characteristic scale at which the random fluctuations or noise in the polarization is suppressed. We call this the polarized correlation length of the visibility, $\lambda_{\widetilde{x}}$, where $x$ is one of the Stokes parameters.

\begin{figure*}
	\includegraphics[trim = 0cm 0cm 0cm 1cm, clip=true,width=\columnwidth]{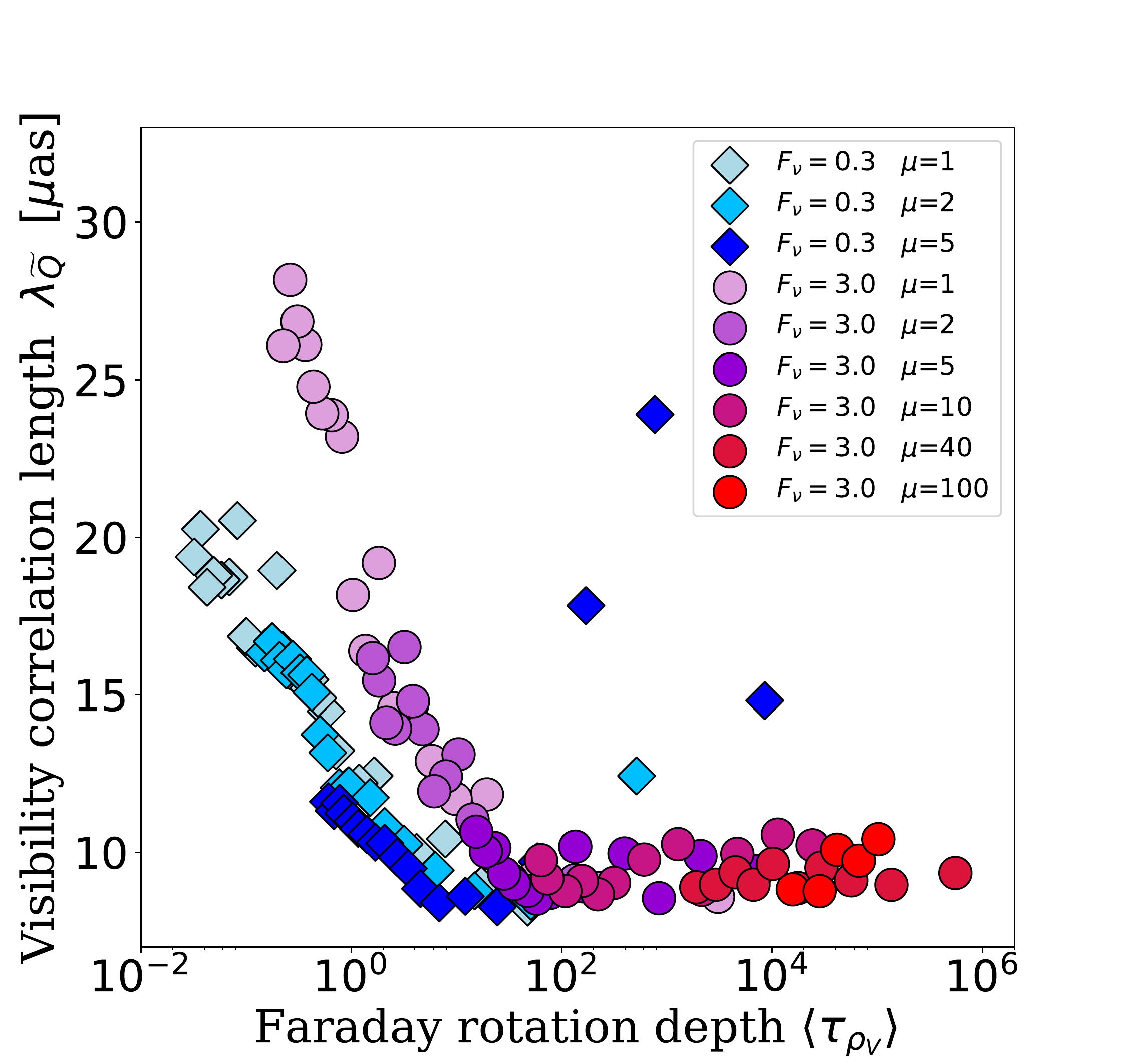}
	\includegraphics[trim = 0cm 0cm 0cm 1cm, clip=true,width=\columnwidth]{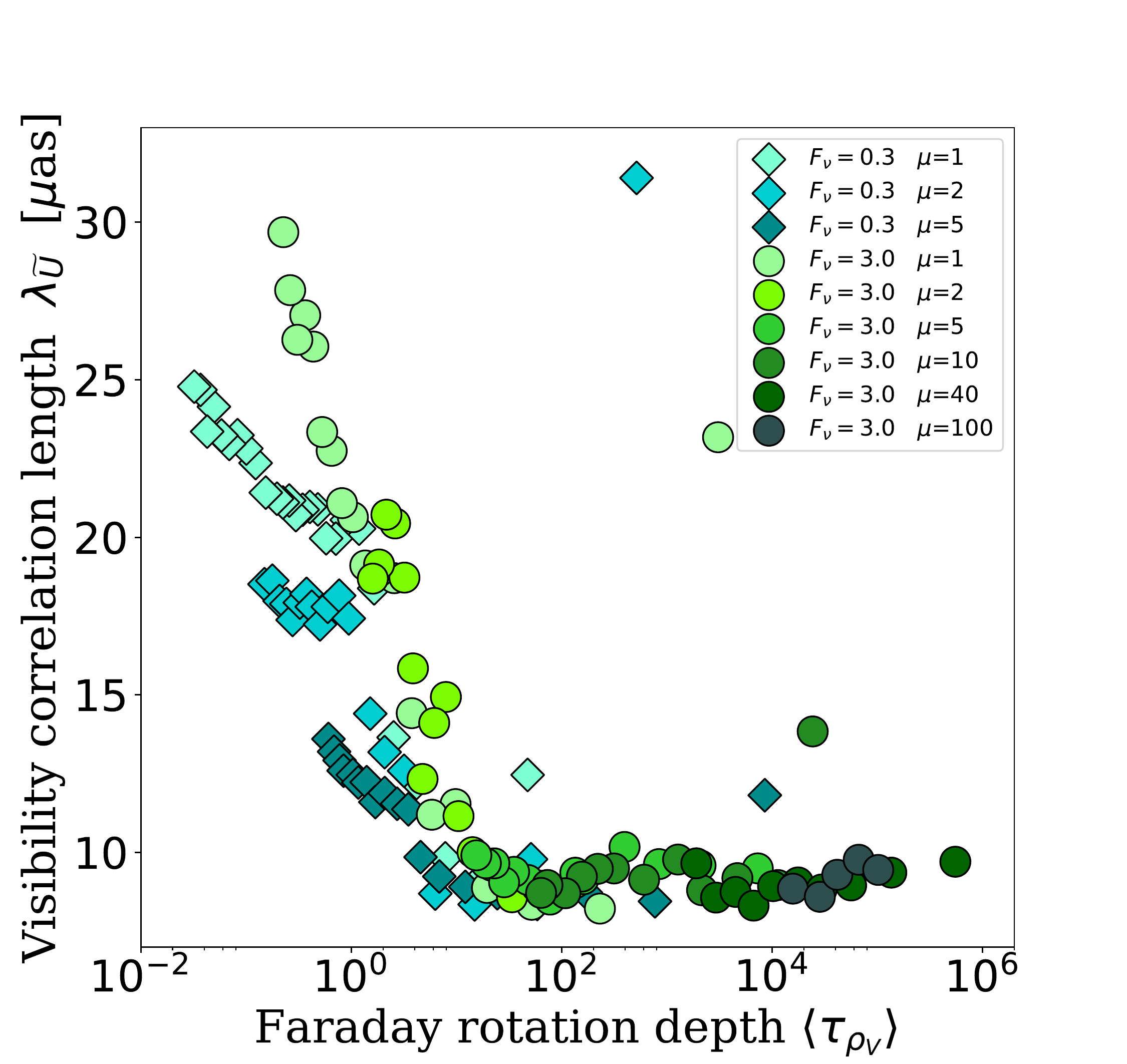}
    \caption{Polarized correlation length in the Fourier domain using
      $\widetilde{Q}$ (left panel) and $\widetilde{U}$ (right panel)
      for our models. The criteria used was to take the inverse of the
      averaged $uv$ distances at which the amplitude of the visibility
      drops below $10\%$ of each respective visibility maximum,
      $\widetilde{Q}_{\rm{max}}$ and $\widetilde{U}_{\rm{max}}$ for
      each model. As in the image domain, the correlation length drops
    sharply for $\langle\tau_{\rho_V} \rangle\gtrsim 1$.}
    \label{fig:corr_length_vis}
\end{figure*}

We measure this as the $uv$ distance at which the visibility's amplitude drops permanently below a certain value. As an example, we have chosen this quantity to be $10\%$ of $\widetilde{Q}_{\rm{max}}$ and $\widetilde{U}_{\rm{max}}$, where $\widetilde{Q}_{\rm{max}}$ and $\widetilde{U}_{\rm{max}}$ are the maximum visibility amplitudes. We define the visibility correlation length as the inverse of the averaged distances which satisfy this criteria. The calculated $\lambda_{\widetilde{Q}}$ and $\lambda_{\widetilde{U}}$ for our models are shown in Fig. \ref{fig:corr_length_vis}. 

As shown in Fig. \ref{fig:corr_length_vis}, the correlation length measured in the visibility space can also
constrain $\langle \tau_{\rho _V} \rangle$ and is measured directly from VLBI observables.
With upcoming data from the EHT, this quantity may be promising for inferring the characteristics of the plasma in the system.
%

\section{Discussion}
\label{sec:discussion}

Sgr A* is a great laboratory for testing accretion physics and general relativity. 
Polarization is a powerful tool for determining the plasma properties and the magnetic field structure.

From a GRMHD simulation of a torus of magnetised plasma in initial hydrostatic equilibrium with a poloidal magnetic field, we have done self-consistent fully relativistic ray tracing radiative transfer calculations of the radiation at $230$ GHz. We have analysed the different polarized images and characterised the degree of coherence in the polarization map as a function of the Faraday rotation depth. This coherence scale we call the correlation length. Large values of this quantity are expected when the Faraday effects are weak and the maps are ordered. Small values of the correlation length in our models point to large Faraday rotation depth values and disordered maps. 

We have proposed a method to relate the polarized correlation length calculated from the images to direct observables of VLBI by taking the Fourier Transform of the images and analysing the large scale structure of the visibilities in the Fourier domain. This shows a similar behaviour to that showed in the image space, with the advantage that it uses VLBI observables.

In the past, unresolved polarization of Sgr A* has been very helpful in constraining models. With the new 
EHT measurements this can be done with the polarization map itself for the first time through the 
correlation length.
So far, the behaviour of this new quantity appears to be model independent, which makes it a promising 
approach that can be used to set restrictions on the plasma parameters around the black hole and distinguish 
models robustly in a way that is often difficult with total intensity images alone.

Constraining $\langle \tau_{\rho _V} \rangle\sim n_{\rm{e}} B T_{\rm{e}}^{-2}$ places limits on the physical properties
of the accreting gas, most directly $T_{\rm{e}}$. In addition, $B^2 \sim \beta^{-1} n T_{\rm{p}}$. From hydrostatic equilibrium, $T_{\rm{p}} \sim T_{\rm vir}
(H/R)^2$, where $T_{\rm vir} \sim m_{\rm{p}} c^2 / r$ is the virial
temperature at dimensionless radius $r = R/R_{\rm s}$ and $H/R$ is the
scale height of the accretion flow. The relative field strength then scales as
$B^2 / n \sim \beta^{-1} (H/R)^2$. At fixed flux density, a limit on
$\langle \tau_{\rho _V} \rangle$ constrains a combination of the magnetic field strength and disc scale height as well as the electron temperature.

We have only considered one inclination. We expect that the trend of
decreasing LP and correlation length with increasing Faraday rotation
optical depth holds at all viewing geometries, but their maximum
values at low Faraday rotation depth will be model-dependent. 

We have also neglected the effects of interstellar scattering. The
diffusive part of the scattering should not affect the correlation
length results, since we use ratios of
the Stokes parameters which are all modified in the same way. The
refractive part of the scattering \citep[e.g.,][]{Gwinn+2014}
could in particular complicate our proposed method for measuring the
correlation length in the Fourier domain, since it will introduce
signal beyond that corresponding to small scale structure in the polarization map.

We have demonstrated the technique with a single snapshot from an axisymmetric GRMHD simulation. 
This is has some limitations given that the MRI is unsustainable in 2D and the simulation can only be
studied for short times.
Therefore, the degree of order seen for
$\langle \tau_{\rho _V} \rangle < 1$ (bottom left panel of Fig. \ref{fig:PM}) is somewhat
overestimated compared to 3D simulations. Extensions to 3D and studying time
variability are goals for future work. The time variable
polarized correlation length could for example be used to measure the
properties of MRI turbulence in EHT data.

We have focused here on the case of mm-VLBI of Sgr A*, but the same
technique should apply to M87 \citep{Moscibrodzka+2017} or any other
synchrotron source with a resolved polarization map. In particular, in
polarized VLBI images of radio jets past work has focused on measuring
the Faraday rotation across the image
\citep[e.g.,][]{Zavala+2003,O'Sullivan+2018}. Here we have shown that
the correlation length may be a more robust indicator of the Faraday
optical depth, if there is a significant contribution from within the
emission region.

\section*{Acknowledgements}
The authors thank M. Johnson, M. Mo\'scibrodzka, C. Gammie, A. Broderick, R. Gold, J. Kim, and
D. P. Marrone for useful discussions related to Sgr A* polarization
and radiative transfer. This work was supported by a CONACyT/DAAD
grant (57265507) and by a Sofja Kovalevskaja 
award from the Alexander von Humboldt foundation.




\bibliographystyle{mnras}
\bibliography{references} 

\begin{thebibliography}{}
\makeatletter
\relax
\def\mn@urlcharsother{\let\do\@makeother \do\$\do\&\do\#\do\^\do\_\do\%\do\~}
\def\mn@doi{\begingroup\mn@urlcharsother \@ifnextchar [ {\mn@doi@}
  {\mn@doi@[]}}
\def\mn@doi@[#1]#2{\def\@tempa{#1}\ifx\@tempa\@empty \href
  {http://dx.doi.org/#2} {doi:#2}\else \href {http://dx.doi.org/#2} {#1}\fi
  \endgroup}
\def\mn@eprint#1#2{\mn@eprint@#1:#2::\@nil}
\def\mn@eprint@arXiv#1{\href {http://arxiv.org/abs/#1} {{\tt arXiv:#1}}}
\def\mn@eprint@dblp#1{\href {http://dblp.uni-trier.de/rec/bibtex/#1.xml}
  {dblp:#1}}
\def\mn@eprint@#1:#2:#3:#4\@nil{\def\@tempa {#1}\def\@tempb {#2}\def\@tempc
  {#3}\ifx \@tempc \@empty \let \@tempc \@tempb \let \@tempb \@tempa \fi \ifx
  \@tempb \@empty \def\@tempb {arXiv}\fi \@ifundefined
  {mn@eprint@\@tempb}{\@tempb:\@tempc}{\expandafter \expandafter \csname
  mn@eprint@\@tempb\endcsname \expandafter{\@tempc}}}

\bibitem[\protect\citeauthoryear{{Agol}}{{Agol}}{2000}]{Agol2000}
{Agol} E.,  2000, \mn@doi [\apjl] {10.1086/312818}, \href
  {http://adsabs.harvard.edu/abs/2000ApJ...538L.121A} {538, L121}

\bibitem[\protect\citeauthoryear{{Aitken}, {Greaves}, {Chrysostomou},
  {Jenness}, {Holland}, {Hough}, {Pierce-Price}  \& {Richer}}{{Aitken}
  et~al.}{2000}]{Aitken+2000}
{Aitken} D.~K.,  {Greaves} J.,  {Chrysostomou} A.,  {Jenness} T.,  {Holland}
  W.,  {Hough} J.~H.,  {Pierce-Price} D.,   {Richer} J.,  2000, \mn@doi [\apjl]
  {10.1086/312685}, \href {http://adsabs.harvard.edu/abs/2000ApJ...534L.173A}
  {534, L173}

\bibitem[\protect\citeauthoryear{{Baganoff} et~al.,}{{Baganoff}
  et~al.}{2001}]{Baganoff+2001}
{Baganoff} F.~K.,  et~al., 2001, \mn@doi [\nat] {10.1038/35092510}, \href
  {http://adsabs.harvard.edu/abs/2001Natur.413...45B} {413, 45}

\bibitem[\protect\citeauthoryear{{Balbus} \& {Hawley}}{{Balbus} \&
  {Hawley}}{1991}]{Balbus+1991}
{Balbus} S.~A.,  {Hawley} J.~F.,  1991, \mn@doi [\apj] {10.1086/170270}, \href
  {http://adsabs.harvard.edu/abs/1991ApJ...376..214B} {376, 214}

\bibitem[\protect\citeauthoryear{{Bardeen}, {Press}  \& {Teukolsky}}{{Bardeen}
  et~al.}{1972}]{Bardeen+1972}
{Bardeen} J.~M.,  {Press} W.~H.,   {Teukolsky} S.~A.,  1972, \mn@doi [\apj]
  {10.1086/151796}, \href {http://adsabs.harvard.edu/abs/1972ApJ...178..347B}
  {178, 347}

\bibitem[\protect\citeauthoryear{{Boehle} et~al.,}{{Boehle}
  et~al.}{2016}]{Boehle+2016}
{Boehle} A.,  et~al., 2016, \mn@doi [\apj] {10.3847/0004-637X/830/1/17}, \href
  {http://adsabs.harvard.edu/abs/2016ApJ...830...17B} {830, 17}

\bibitem[\protect\citeauthoryear{{Bower}, {Wright}, {Falcke}  \&
  {Backer}}{{Bower} et~al.}{2003}]{Bower+2003}
{Bower} G.~C.,  {Wright} M.~C.~H.,  {Falcke} H.,   {Backer} D.~C.,  2003,
  \mn@doi [\apj] {10.1086/373989}, \href
  {http://adsabs.harvard.edu/abs/2003ApJ...588..331B} {588, 331}

\bibitem[\protect\citeauthoryear{{Bower} et~al.,}{{Bower}
  et~al.}{2015}]{Bower+2015}
{Bower} G.~C.,  et~al., 2015, \mn@doi [\apj] {10.1088/0004-637X/802/1/69},
  \href {http://adsabs.harvard.edu/abs/2015ApJ...802...69B} {802, 69}

\bibitem[\protect\citeauthoryear{{Broderick} \& {Loeb}}{{Broderick} \&
  {Loeb}}{2006}]{Broderick+2006}
{Broderick} A.~E.,  {Loeb} A.,  2006, \mn@doi [\mnras]
  {10.1111/j.1365-2966.2006.10152.x}, \href
  {http://adsabs.harvard.edu/abs/2006MNRAS.367..905B} {367, 905}

\bibitem[\protect\citeauthoryear{{Broderick}, {Fish}, {Doeleman}  \&
  {Loeb}}{{Broderick} et~al.}{2011}]{Broderick+2011}
{Broderick} A.~E.,  {Fish} V.~L.,  {Doeleman} S.~S.,   {Loeb} A.,  2011,
  \mn@doi [\apj] {10.1088/0004-637X/735/2/110}, \href
  {http://adsabs.harvard.edu/abs/2011ApJ...735..110B} {735, 110}

\bibitem[\protect\citeauthoryear{{Bromley}, {Melia}  \& {Liu}}{{Bromley}
  et~al.}{2001}]{Bromley+2001}
{Bromley} B.~C.,  {Melia} F.,   {Liu} S.,  2001, \mn@doi [\apjl]
  {10.1086/322862}, \href {http://adsabs.harvard.edu/abs/2001ApJ...555L..83B}
  {555, L83}

\bibitem[\protect\citeauthoryear{{Chael}, {Rowan}, {Narayan}, {Johnson}  \&
  {Sironi}}{{Chael} et~al.}{2018}]{Chael+2018}
{Chael} A.,  {Rowan} M.~E.,  {Narayan} R.,  {Johnson} M.~D.,   {Sironi} L.,
  2018, preprint, \href {http://adsabs.harvard.edu/abs/2018arXiv180406416C} {}
  (\mn@eprint {arXiv} {1804.06416})

\bibitem[\protect\citeauthoryear{{Chan}, {Psaltis}, {{\"O}zel}, {Narayan}  \&
  {Sa{\c d}owski}}{{Chan} et~al.}{2015}]{Chan+2015}
{Chan} C.-K.,  {Psaltis} D.,  {{\"O}zel} F.,  {Narayan} R.,   {Sa{\c d}owski}
  A.,  2015, \mn@doi [\apj] {10.1088/0004-637X/799/1/1}, \href
  {http://adsabs.harvard.edu/abs/2015ApJ...799....1C} {799, 1}

\bibitem[\protect\citeauthoryear{{Dexter}}{{Dexter}}{2016}]{Dexter2016}
{Dexter} J.,  2016, \mn@doi [\mnras] {10.1093/mnras/stw1526}, \href
  {http://adsabs.harvard.edu/abs/2016MNRAS.462..115D} {462, 115}

\bibitem[\protect\citeauthoryear{{Dexter} \& {Agol}}{{Dexter} \&
  {Agol}}{2009}]{Dexter+2009}
{Dexter} J.,  {Agol} E.,  2009, \mn@doi [\apj] {10.1088/0004-637X/696/2/1616},
  \href {http://adsabs.harvard.edu/abs/2009ApJ...696.1616D} {696, 1616}

\bibitem[\protect\citeauthoryear{{Dexter}, {Agol}, {Fragile}  \&
  {McKinney}}{{Dexter} et~al.}{2010}]{Dexter+2010}
{Dexter} J.,  {Agol} E.,  {Fragile} P.~C.,   {McKinney} J.~C.,  2010, \mn@doi
  [\apj] {10.1088/0004-637X/717/2/1092}, \href
  {http://adsabs.harvard.edu/abs/2010ApJ...717.1092D} {717, 1092}

\bibitem[\protect\citeauthoryear{{Doeleman} et~al.,}{{Doeleman}
  et~al.}{2008}]{Doeleman+2008}
{Doeleman} S.~S.,  et~al., 2008, \mn@doi [\nat] {10.1038/nature07245}, \href
  {http://adsabs.harvard.edu/abs/2008Natur.455...78D} {455, 78}

\bibitem[\protect\citeauthoryear{{Falcke} \& {Markoff}}{{Falcke} \&
  {Markoff}}{2000}]{Falcke+2000}
{Falcke} H.,  {Markoff} S.,  2000, \aap, \href
  {http://adsabs.harvard.edu/abs/2000A%26A...362..113F} {362, 113}

\bibitem[\protect\citeauthoryear{{Falcke} \& {Markoff}}{{Falcke} \&
  {Markoff}}{2013}]{Falcke+2013}
{Falcke} H.,  {Markoff} S.~B.,  2013, \mn@doi [Classical and Quantum Gravity]
  {10.1088/0264-9381/30/24/244003}, \href
  {http://adsabs.harvard.edu/abs/2013CQGra..30x4003F} {30, 244003}

\bibitem[\protect\citeauthoryear{{Fish} et~al.,}{{Fish}
  et~al.}{2011}]{Fish+2011}
{Fish} V.~L.,  et~al., 2011, \mn@doi [\apjl] {10.1088/2041-8205/727/2/L36},
  \href {http://adsabs.harvard.edu/abs/2011ApJ...727L..36F} {727, L36}

\bibitem[\protect\citeauthoryear{{Fishbone} \& {Moncrief}}{{Fishbone} \&
  {Moncrief}}{1976}]{Fishbone+1976}
{Fishbone} L.~G.,  {Moncrief} V.,  1976, \mn@doi [\apj] {10.1086/154565}, \href
  {http://adsabs.harvard.edu/abs/1976ApJ...207..962F} {207, 962}

\bibitem[\protect\citeauthoryear{{Gammie}, {McKinney}  \& {T{\'o}th}}{{Gammie}
  et~al.}{2003}]{Gammie+2003}
{Gammie} C.~F.,  {McKinney} J.~C.,   {T{\'o}th} G.,  2003, \mn@doi [\apj]
  {10.1086/374594}, \href {http://adsabs.harvard.edu/abs/2003ApJ...589..444G}
  {589, 444}

\bibitem[\protect\citeauthoryear{{Genzel}, {Eisenhauer}  \&
  {Gillessen}}{{Genzel} et~al.}{2010}]{Genzel+2010}
{Genzel} R.,  {Eisenhauer} F.,   {Gillessen} S.,  2010, \mn@doi [Reviews of
  Modern Physics] {10.1103/RevModPhys.82.3121}, \href
  {http://adsabs.harvard.edu/abs/2010RvMP...82.3121G} {82, 3121}

\bibitem[\protect\citeauthoryear{{Gillessen} et~al.,}{{Gillessen}
  et~al.}{2017}]{Gillessen+2017}
{Gillessen} S.,  et~al., 2017, \mn@doi [\apj] {10.3847/1538-4357/aa5c41}, \href
  {http://adsabs.harvard.edu/abs/2017ApJ...837...30G} {837, 30}

\bibitem[\protect\citeauthoryear{{Gold}, {McKinney}, {Johnson}  \&
  {Doeleman}}{{Gold} et~al.}{2017}]{Gold+2017}
{Gold} R.,  {McKinney} J.~C.,  {Johnson} M.~D.,   {Doeleman} S.~S.,  2017,
  \mn@doi [\apj] {10.3847/1538-4357/aa6193}, \href
  {http://adsabs.harvard.edu/abs/2017ApJ...837..180G} {837, 180}

\bibitem[\protect\citeauthoryear{{Gwinn}, {Kovalev}, {Johnson}  \&
  {Soglasnov}}{{Gwinn} et~al.}{2014}]{Gwinn+2014}
{Gwinn} C.~R.,  {Kovalev} Y.~Y.,  {Johnson} M.~D.,   {Soglasnov} V.~A.,  2014,
  \mn@doi [\apjl] {10.1088/2041-8205/794/1/L14}, \href
  {http://adsabs.harvard.edu/abs/2014ApJ...794L..14G} {794, L14}

\bibitem[\protect\citeauthoryear{{Howes}}{{Howes}}{2010}]{Howes2010}
{Howes} G.~G.,  2010, \mn@doi [\mnras] {10.1111/j.1745-3933.2010.00958.x},
  \href {http://adsabs.harvard.edu/abs/2010MNRAS.409L.104H} {409, L104}

\bibitem[\protect\citeauthoryear{{Johnson} et~al.,}{{Johnson}
  et~al.}{2015}]{Johnson+2015}
{Johnson} M.~D.,  et~al., 2015, \mn@doi [Science] {10.1126/science.aac7087},
  \href {http://adsabs.harvard.edu/abs/2015Sci...350.1242J} {350, 1242}

\bibitem[\protect\citeauthoryear{{Jones} \& {Hardee}}{{Jones} \&
  {Hardee}}{1979}]{Jones+1979}
{Jones} T.~W.,  {Hardee} P.~E.,  1979, \mn@doi [\apj] {10.1086/156843}, \href
  {http://adsabs.harvard.edu/abs/1979ApJ...228..268J} {228, 268}

\bibitem[\protect\citeauthoryear{{Kamruddin} \& {Dexter}}{{Kamruddin} \&
  {Dexter}}{2013}]{Kamruddin+2013}
{Kamruddin} A.~B.,  {Dexter} J.,  2013, \mn@doi [\mnras]
  {10.1093/mnras/stt1068}, \href
  {http://adsabs.harvard.edu/abs/2013MNRAS.434..765K} {434, 765}

\bibitem[\protect\citeauthoryear{{Marrone}, {Moran}, {Zhao}  \&
  {Rao}}{{Marrone} et~al.}{2006}]{Marrone+2006}
{Marrone} D.~P.,  {Moran} J.~M.,  {Zhao} J.-H.,   {Rao} R.,  2006, \mn@doi
  [\apj] {10.1086/500106}, \href
  {http://adsabs.harvard.edu/abs/2006ApJ...640..308M} {640, 308}

\bibitem[\protect\citeauthoryear{{McKinney}}{{McKinney}}{2006}]{McKinney2006}
{McKinney} J.~C.,  2006, \mn@doi [\mnras] {10.1111/j.1365-2966.2006.10256.x},
  \href {http://adsabs.harvard.edu/abs/2006MNRAS.368.1561M} {368, 1561}

\bibitem[\protect\citeauthoryear{{Mo{\'s}cibrodzka}, {Gammie}, {Dolence},
  {Shiokawa}  \& {Leung}}{{Mo{\'s}cibrodzka} et~al.}{2009}]{Moscibrodzka+2009}
{Mo{\'s}cibrodzka} M.,  {Gammie} C.~F.,  {Dolence} J.~C.,  {Shiokawa} H.,
  {Leung} P.~K.,  2009, \mn@doi [\apj] {10.1088/0004-637X/706/1/497}, \href
  {http://adsabs.harvard.edu/abs/2009ApJ...706..497M} {706, 497}

\bibitem[\protect\citeauthoryear{{Mo{\'s}cibrodzka}, {Falcke}, {Shiokawa}  \&
  {Gammie}}{{Mo{\'s}cibrodzka} et~al.}{2014}]{Moscibrodzka+2014}
{Mo{\'s}cibrodzka} M.,  {Falcke} H.,  {Shiokawa} H.,   {Gammie} C.~F.,  2014,
  \mn@doi [\aap] {10.1051/0004-6361/201424358}, \href
  {http://adsabs.harvard.edu/abs/2014A%26A...570A...7M} {570, A7}

\bibitem[\protect\citeauthoryear{{Mo{\'s}cibrodzka}, {Falcke}  \&
  {Shiokawa}}{{Mo{\'s}cibrodzka} et~al.}{2016}]{Moscibrodzka+2016}
{Mo{\'s}cibrodzka} M.,  {Falcke} H.,   {Shiokawa} H.,  2016, \mn@doi [\aap]
  {10.1051/0004-6361/201526630}, \href
  {http://adsabs.harvard.edu/abs/2016A%26A...586A..38M} {586, A38}

\bibitem[\protect\citeauthoryear{{Mo{\'s}cibrodzka}, {Dexter}, {Davelaar}  \&
  {Falcke}}{{Mo{\'s}cibrodzka} et~al.}{2017}]{Moscibrodzka+2017}
{Mo{\'s}cibrodzka} M.,  {Dexter} J.,  {Davelaar} J.,   {Falcke} H.,  2017,
  \mn@doi [\mnras] {10.1093/mnras/stx587}, \href
  {http://adsabs.harvard.edu/abs/2017MNRAS.468.2214M} {468, 2214}

\bibitem[\protect\citeauthoryear{{Narayan} \& {Yi}}{{Narayan} \&
  {Yi}}{1995}]{Narayan+1995}
{Narayan} R.,  {Yi} I.,  1995, \mn@doi [\apj] {10.1086/176343}, \href
  {http://adsabs.harvard.edu/abs/1995ApJ...452..710N} {452, 710}

\bibitem[\protect\citeauthoryear{{Narayan}, {S{\"A} dowski}, {Penna}  \&
  {Kulkarni}}{{Narayan} et~al.}{2012}]{Narayan+2012}
{Narayan} R.,  {S{\"A} dowski} A.,  {Penna} R.~F.,   {Kulkarni} A.~K.,  2012,
  \mn@doi [\mnras] {10.1111/j.1365-2966.2012.22002.x}, \href
  {http://adsabs.harvard.edu/abs/2012MNRAS.426.3241N} {426, 3241}

\bibitem[\protect\citeauthoryear{{Noble}, {Gammie}, {McKinney}  \& {Del
  Zanna}}{{Noble} et~al.}{2006}]{Noble+2006}
{Noble} S.~C.,  {Gammie} C.~F.,  {McKinney} J.~C.,   {Del Zanna} L.,  2006,
  \mn@doi [\apj] {10.1086/500349}, \href
  {http://adsabs.harvard.edu/abs/2006ApJ...641..626N} {641, 626}

\bibitem[\protect\citeauthoryear{{O'Sullivan}, {Lenc}, {Anderson}, {Gaensler}
  \& {Murphy}}{{O'Sullivan} et~al.}{2018}]{O'Sullivan+2018}
{O'Sullivan} S.~P.,  {Lenc} E.,  {Anderson} C.~S.,  {Gaensler} B.~M.,
  {Murphy} T.,  2018, \mn@doi [\mnras] {10.1093/mnras/sty171}, \href
  {http://adsabs.harvard.edu/abs/2018MNRAS.475.4263O} {475, 4263}

\bibitem[\protect\citeauthoryear{{Quataert} \& {Gruzinov}}{{Quataert} \&
  {Gruzinov}}{2000}]{Quataert+2000}
{Quataert} E.,  {Gruzinov} A.,  2000, \mn@doi [\apj] {10.1086/317845}, \href
  {http://adsabs.harvard.edu/abs/2000ApJ...545..842Q} {545, 842}

\bibitem[\protect\citeauthoryear{{Quataert} \& {Narayan}}{{Quataert} \&
  {Narayan}}{1999}]{Quataert+1999}
{Quataert} E.,  {Narayan} R.,  1999, \mn@doi [\apj] {10.1086/307097}, \href
  {http://adsabs.harvard.edu/abs/1999ApJ...516..399Q} {516, 399}

\bibitem[\protect\citeauthoryear{{Ressler}, {Tchekhovskoy}, {Quataert},
  {Chandra}  \& {Gammie}}{{Ressler} et~al.}{2015}]{Ressler+2015}
{Ressler} S.~M.,  {Tchekhovskoy} A.,  {Quataert} E.,  {Chandra} M.,   {Gammie}
  C.~F.,  2015, \mn@doi [\mnras] {10.1093/mnras/stv2084}, \href
  {http://adsabs.harvard.edu/abs/2015MNRAS.454.1848R} {454, 1848}

\bibitem[\protect\citeauthoryear{{Ressler}, {Tchekhovskoy}, {Quataert}  \&
  {Gammie}}{{Ressler} et~al.}{2017}]{Ressler+2017}
{Ressler} S.~M.,  {Tchekhovskoy} A.,  {Quataert} E.,   {Gammie} C.~F.,  2017,
  \mn@doi [\mnras] {10.1093/mnras/stx364}, \href
  {http://adsabs.harvard.edu/abs/2017MNRAS.467.3604R} {467, 3604}

\bibitem[\protect\citeauthoryear{{Rowan}, {Sironi}  \& {Narayan}}{{Rowan}
  et~al.}{2017}]{Rowan+2017}
{Rowan} M.~E.,  {Sironi} L.,   {Narayan} R.,  2017, \mn@doi [\apj]
  {10.3847/1538-4357/aa9380}, \href
  {http://adsabs.harvard.edu/abs/2017ApJ...850...29R} {850, 29}

\bibitem[\protect\citeauthoryear{{Shcherbakov} \& {Huang}}{{Shcherbakov} \&
  {Huang}}{2011}]{ShcherbakovHuang2011}
{Shcherbakov} R.~V.,  {Huang} L.,  2011, \mn@doi [\mnras]
  {10.1111/j.1365-2966.2010.17502.x}, \href
  {http://adsabs.harvard.edu/abs/2011MNRAS.410.1052S} {410, 1052}

\bibitem[\protect\citeauthoryear{{Werner}, {Uzdensky}, {Begelman}, {Cerutti}
  \& {Nalewajko}}{{Werner} et~al.}{2018}]{Werner+2018}
{Werner} G.~R.,  {Uzdensky} D.~A.,  {Begelman} M.~C.,  {Cerutti} B.,
  {Nalewajko} K.,  2018, \mn@doi [\mnras] {10.1093/mnras/stx2530}, \href
  {http://adsabs.harvard.edu/abs/2018MNRAS.473.4840W} {473, 4840}

\bibitem[\protect\citeauthoryear{{Yuan}, {Markoff}  \& {Falcke}}{{Yuan}
  et~al.}{2002}]{Yuan+2002}
{Yuan} F.,  {Markoff} S.,   {Falcke} H.,  2002, \mn@doi [\aap]
  {10.1051/0004-6361:20011709}, \href
  {http://adsabs.harvard.edu/abs/2002A%26A...383..854Y} {383, 854}

\bibitem[\protect\citeauthoryear{{Yuan}, {Quataert}  \& {Narayan}}{{Yuan}
  et~al.}{2003}]{Yuan+2003}
{Yuan} F.,  {Quataert} E.,   {Narayan} R.,  2003, \mn@doi [\apj]
  {10.1086/378716}, \href {http://adsabs.harvard.edu/abs/2003ApJ...598..301Y}
  {598, 301}

\bibitem[\protect\citeauthoryear{{Zavala} \& {Taylor}}{{Zavala} \&
  {Taylor}}{2003}]{Zavala+2003}
{Zavala} R.~T.,  {Taylor} G.~B.,  2003, \mn@doi [\apj] {10.1086/374619}, \href
  {http://adsabs.harvard.edu/abs/2003ApJ...589..126Z} {589, 126}

\makeatother
\end{thebibliography}







\label{lastpage}
\end{document}